\documentclass{svproc}
\usepackage{url}

\usepackage{hyperref}
\usepackage{type1cm}
\usepackage{makeidx}
\usepackage{graphicx}
\graphicspath{{Images/}{../12Alias/Images/BRDF/}}

\usepackage{multicol}
\usepackage[bottom]{footmisc}

\usepackage{newtxtext}
\usepackage[varvw]{newtxmath}
\usepackage{bm}
\usepackage{siunitx}
\newcounter{algorithm}

\newcommand{\bsE}{{\boldsymbol{E}}}

\DeclareSymbolFont{bbold}{U}{bbold}{m}{n}
\DeclareSymbolFontAlphabet{\mathbbold}{bbold}

\usepackage{xcolor}
\definecolor{HighlightColor}{rgb}{0.462745098, 0.725490196, 0.000000000}
\hypersetup{colorlinks=true, allcolors=HighlightColor}

\begin{document}

\mainmatter
\title{Parametric Integration with Neural Integral Operators}
\titlerunning{Parametric Integration with Neural Integral Operators}
\author{Christoph Schied\inst{1} \and Alexander Keller\inst{2}}
\authorrunning{Christoph Schied and Alexander Keller}
\institute{\email{cschied@nvidia.com} \and \email{akeller@nvidia.com} \\
NVIDIA, Fasanenstr. 81, 10623 Berlin, Germany}

\maketitle

\begin{abstract}
Real-time rendering imposes strict limitations on the sampling budget
for light transport simulation, often resulting in noisy images. However, denoisers
have demonstrated that it is possible to produce noise-free images through filtering.
We enhance image quality by removing noise before material shading, rather
than filtering already shaded noisy images. This approach allows for
material-agnostic denoising (MAD) and leverages machine learning by
approximating the light transport integral operator with a neural network, effectively performing
parametric integration with neural operators. Our method operates in real-time,
requires data from only a single frame, seamlessly integrates with existing
denoisers and temporal anti-aliasing techniques, and is efficient to train. Additionally, it
is straightforward to incorporate with physically based rendering algorithms.
\end{abstract}

\section{Introduction}

Light transport for photo-realistic image synthesis is efficiently
simulated by accumulating the contributions of sampled light transport
paths that connect cameras to light sources \cite{PBRT}.
However, in real-time rendering, practical compute budgets severely limit the
number of light transport paths, leading to images that may
exhibit noticeable noise due to insufficient sampling.

Interestingly, a human observer easily understands noisy images.
Based on the observation that a noisy image still contains a
sufficient amount of information to recover it from noise, filtering
techniques \cite{ATrous,U-Net,Bako17} have been devised.
While early denoisers are based on classic image processing,
advanced algorithms  build on neural networks.

We present a machine-learning-driven algorithm that filters noise
prior to shading, addressing the issue that post-shading filtering often blurs essential
material details and obscures fine geometric features.  To achieve this,
we project the incident radiance onto a higher-dimensional function space
and train a neural integral operator acting on that space to approximate the original parametric integral.

Projections of the incident radiance are known to be much smoother
than the parametric integral we aim to compute \cite{Krivanek:2009:PGI}. Therefore,
denoising the incident radiance -- effectively filtering the integrand
before parametric integration -- instead of the shaded image,
helps preserve crucial scene features and renders crisper, more
accurate images. By training the integral operator, we eliminate the need for handcrafted heuristics.

\section{Neural Shading Pipeline} \label{Sec:Algorithm}

Given a point $x$ on a surface and a direction $\omega_r$ of observation, we strive to approximate the parametric integral representing the reflected radiance
\begin{eqnarray}
  L_r(x, \omega_r) & = & \int_{{\mathcal S}^2_+(x)} f(x, \omega, \omega_r) L_i(x, \omega) \cos \vartheta \, d \omega  \label{Eqn:IEQ} 
  \qquad \raisebox{-7ex}{\includegraphics[width=0.32\linewidth]{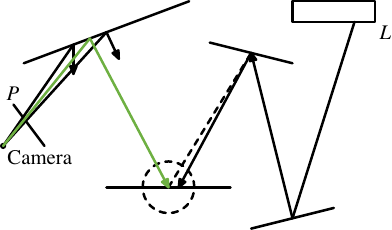}} \\
  & \approx & M_\theta( x, \omega_r, \underbrace{\int_{{\mathcal S}^2_+(x)} \bsE(\omega, \omega_r) L_i(x, \omega) \cos \vartheta \, d \omega}_{\text{projected irradiance } \mu^*} \; ) \label{Eqn:Approximation}
\end{eqnarray}
by a machine learned function $M_\theta$.
The operator acts on the projected irradiance $\mu^*$, which is the integral
that projects the incident radiance $L_i(x, \omega)$ onto the vector-valued function $\bsE(\omega, \omega_r)$.
The unit hemisphere ${\mathcal S}^2_+(x)$ around $x$ is oriented by the surface normal in $x$
and $\vartheta$ is the angle between the surface normal in $x$ and the direction $\omega$. The
integral kernel $f$ is the bidirectional scattering distribution function (BSDF) characterizing the
optical properties of a surface when interacting with light.

\paragraph{Algorithm Overview:} In order to compute an image, light transport paths are sampled for each pixel,
starting with a ray traced from a virtual camera to find the closest intersection $x$
with the scene surface. Simultaneously for all $x$, our neural shading pipeline then approximates the parametric
integral~(\ref{Eqn:IEQ}) in real-time in
three steps: First, the incident radiance is projected onto $\bsE(\omega, \omega_r)$
(Sect.~\ref{Sec:IrradianceEncoder}). Second, the noise is filtered (Sect.~\ref{Sec:Denoiser})
in the space spanned by the vector-valued function $\bsE(\omega, \omega_r)$.
Third, the material decoder $M_\theta$ determines the shading (Sect.~\ref{Sec:MaterialDecoder})
using the filtered irradiance. Exemplary results are shown in Sect.~\ref{Sec:Results}.

The efficient training of $M_\theta$ and appropriate loss functions are
explained in Sect.~\ref{Sec:Training}, followed by a discussion of our algorithm
in Sect.~\ref{Sec:Discussion}.
The details of light transport simulation by path tracing~\cite{PBRT} are beyond the scope of the manuscript,
but most interesting to study. For our context, a compact primer is the prior work by Nalbach et al.~\cite{DeepShading}
on applying neural networks to image synthesis.

\begin{figure}
  \centering
  \begin{tabular}{lccc}
  \raisebox{6ex}{$E_0$} & \includegraphics[width=0.31\linewidth]{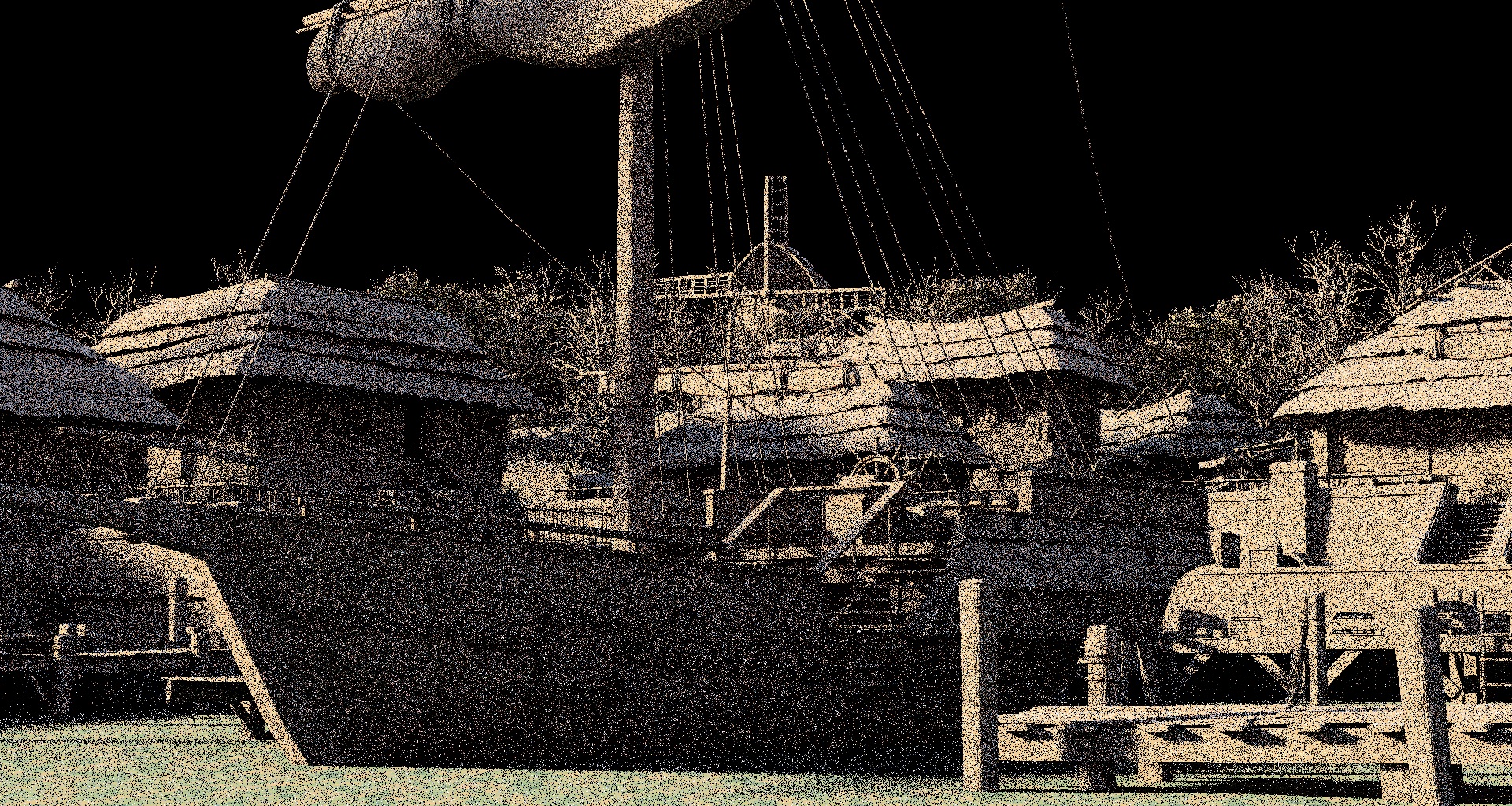}
  & \includegraphics[width=0.31\linewidth]{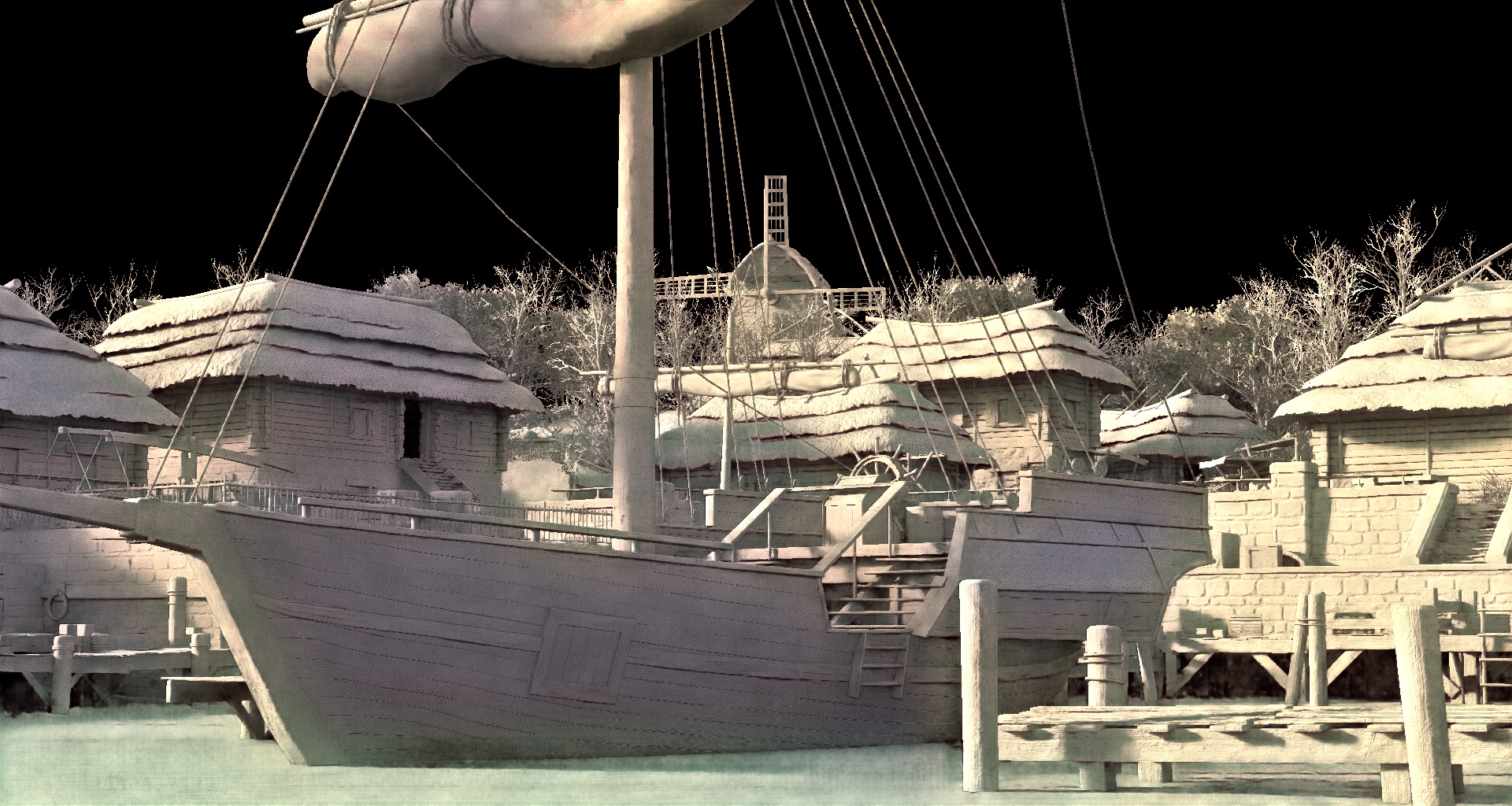}
  & \includegraphics[width=0.31\linewidth]{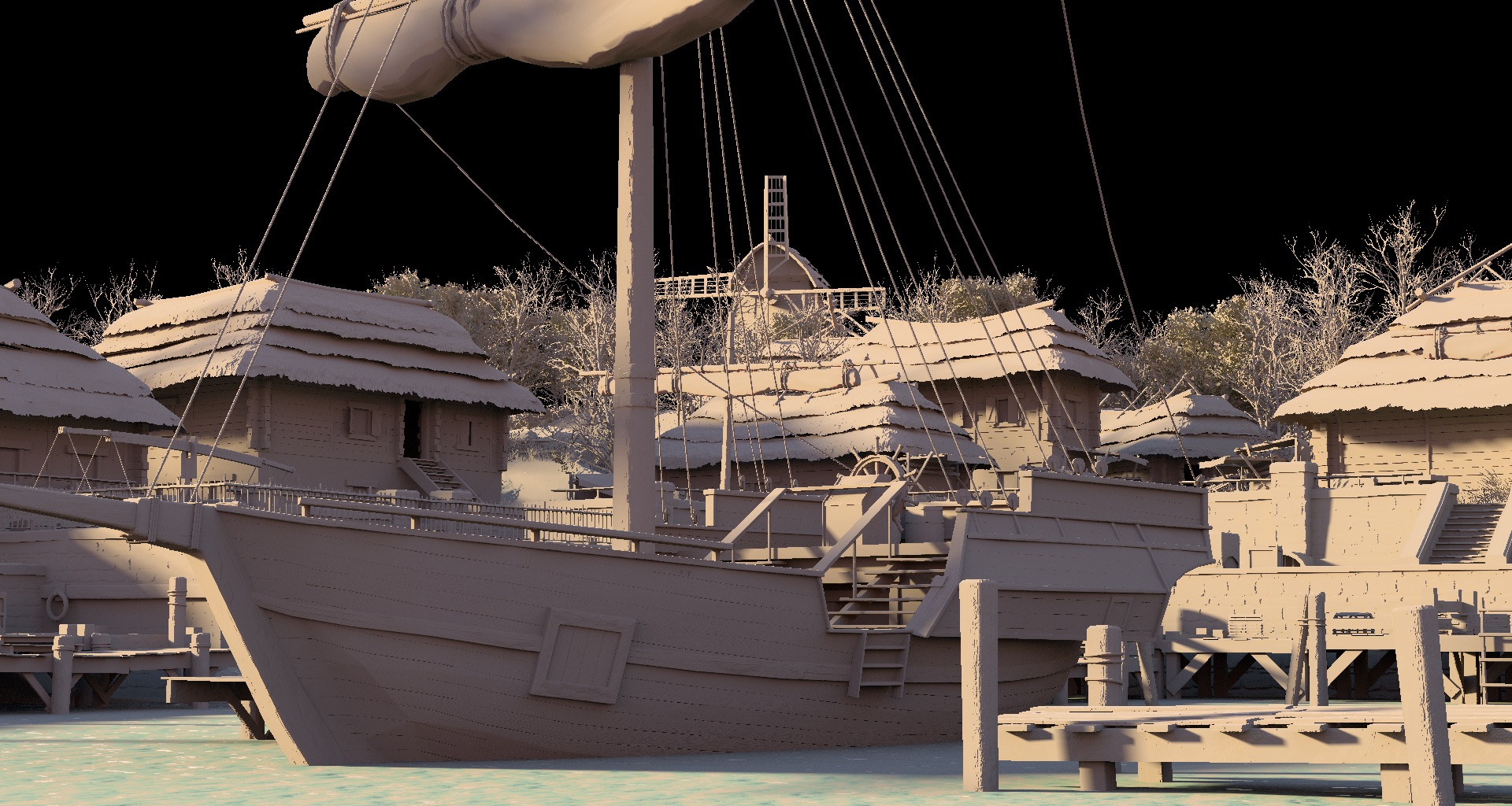}
  \\
  \raisebox{6ex}{$E_1$} & \includegraphics[width=0.31\linewidth]{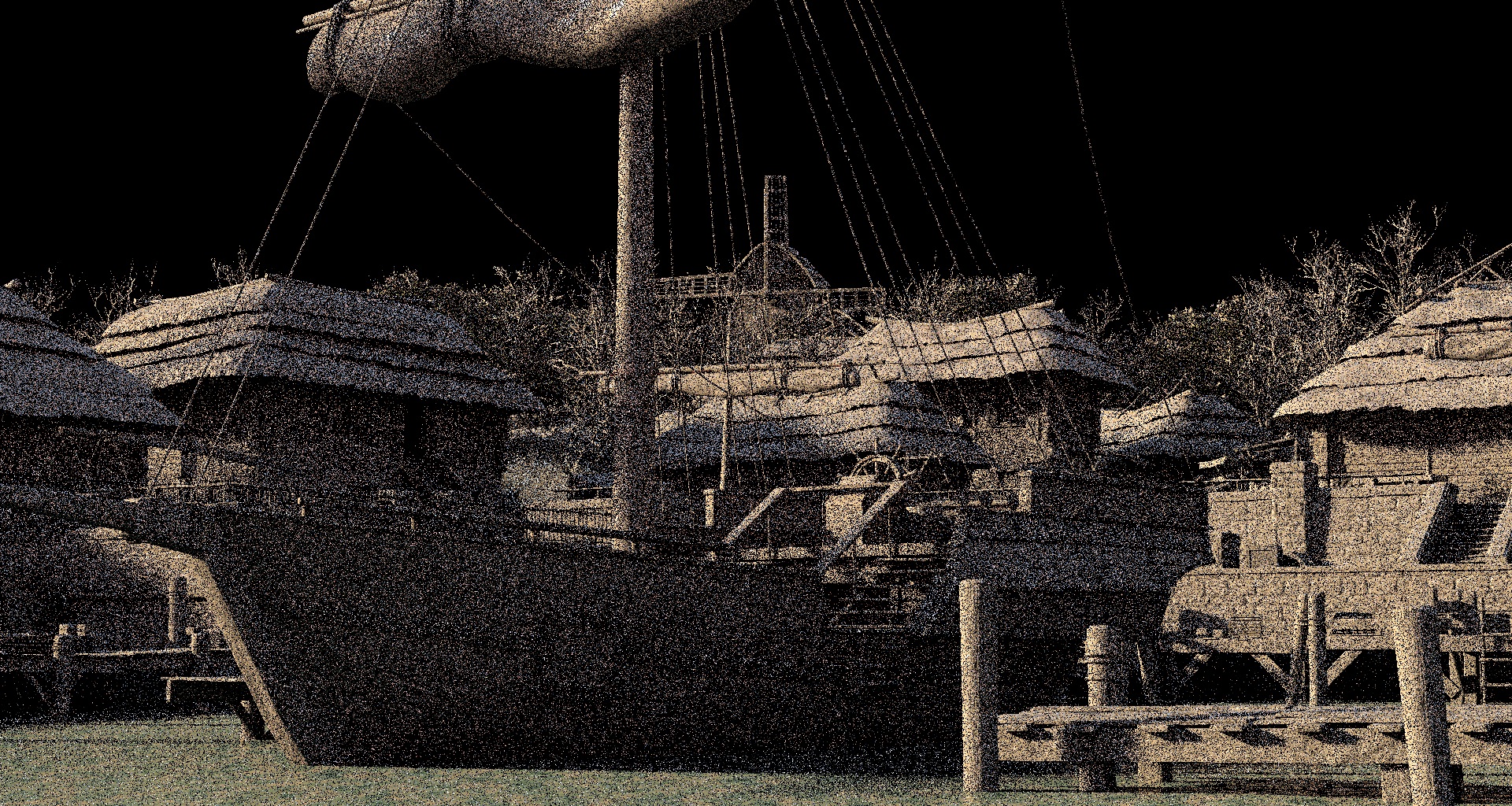}
  & \includegraphics[width=0.31\linewidth]{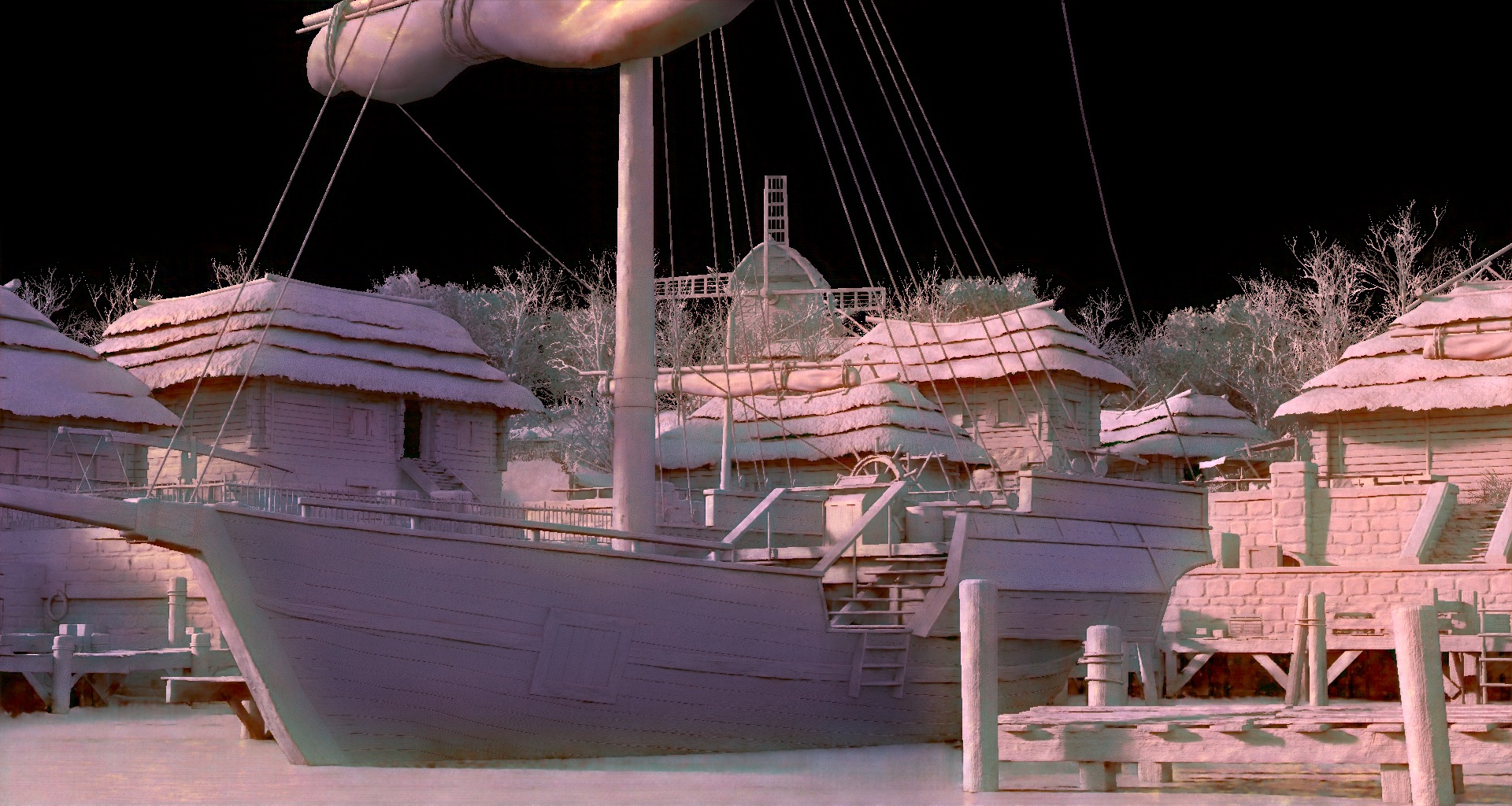}
  & \includegraphics[width=0.31\linewidth]{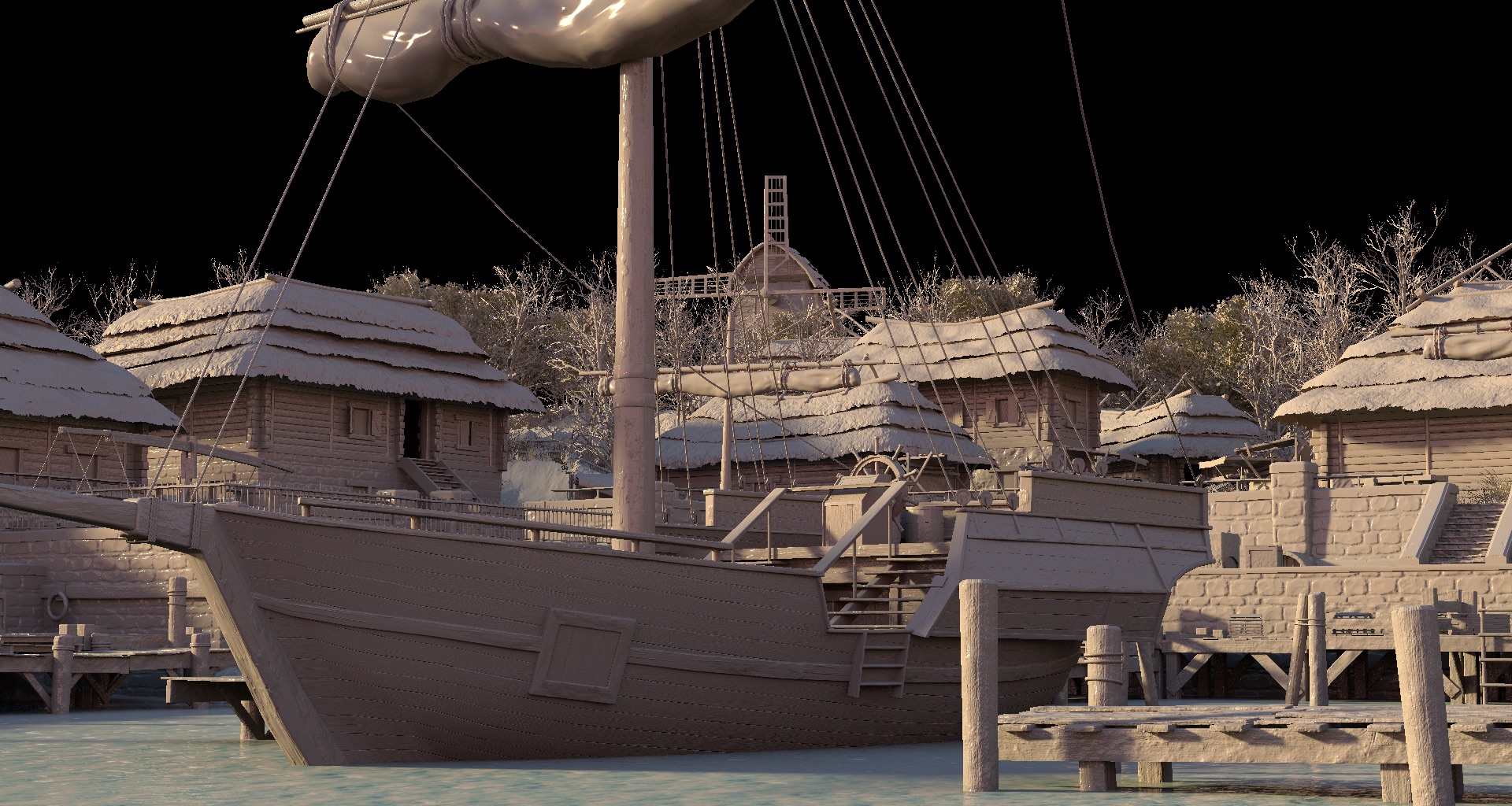}
  \\
  \raisebox{6ex}{$E_2$} & \includegraphics[width=0.31\linewidth]{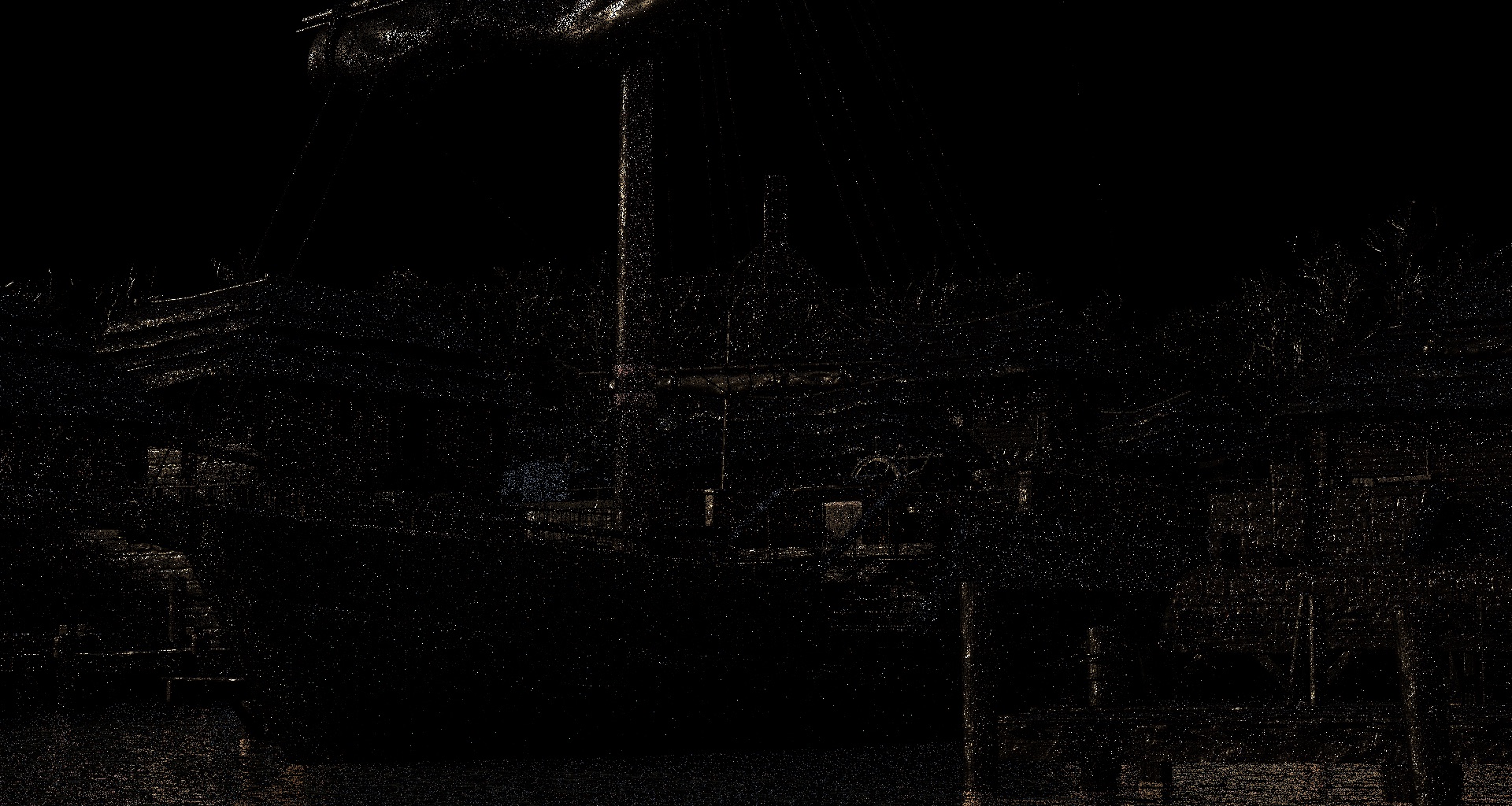}
  & \includegraphics[width=0.31\linewidth]{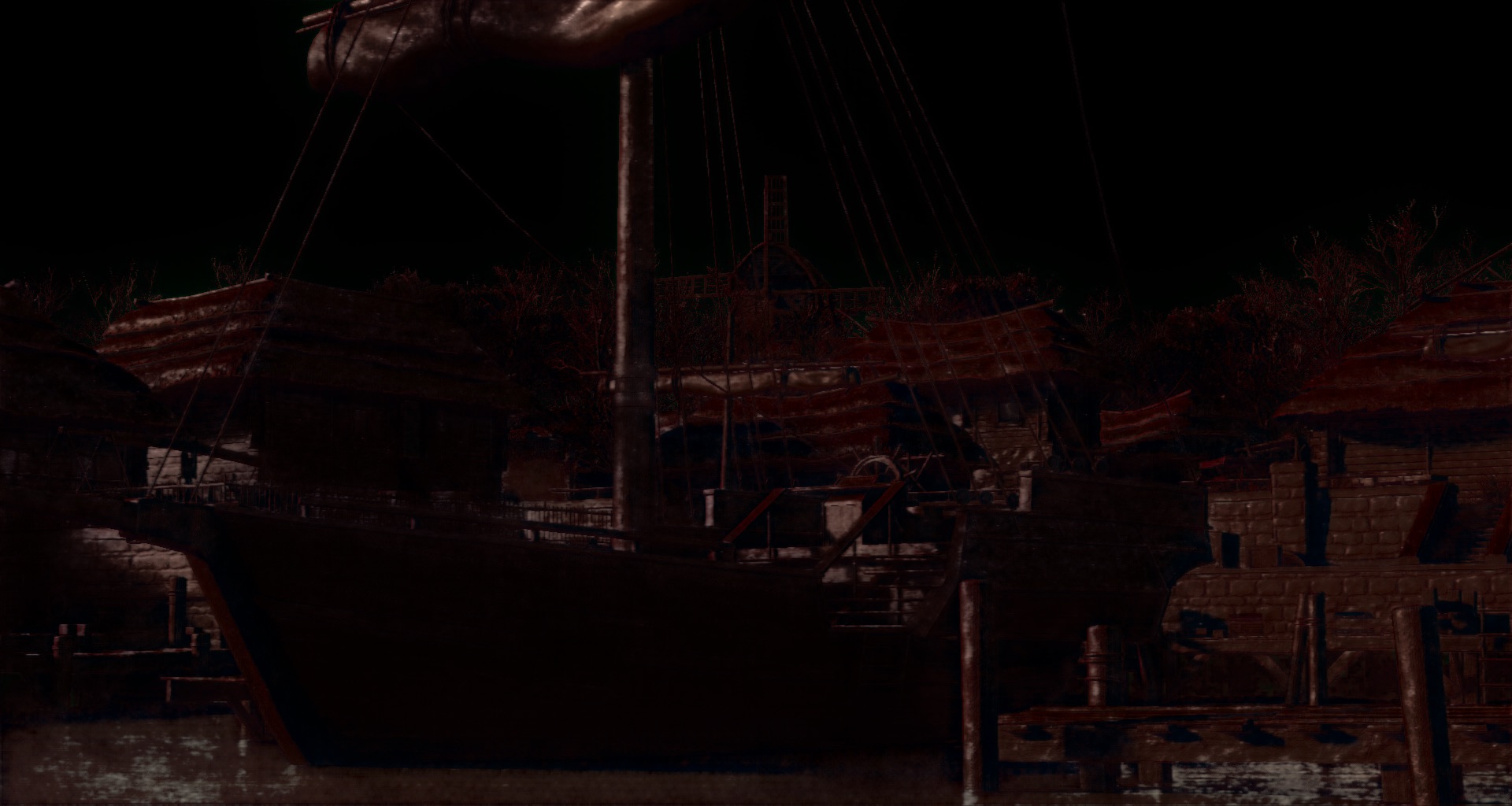}
  & \includegraphics[width=0.31\linewidth]{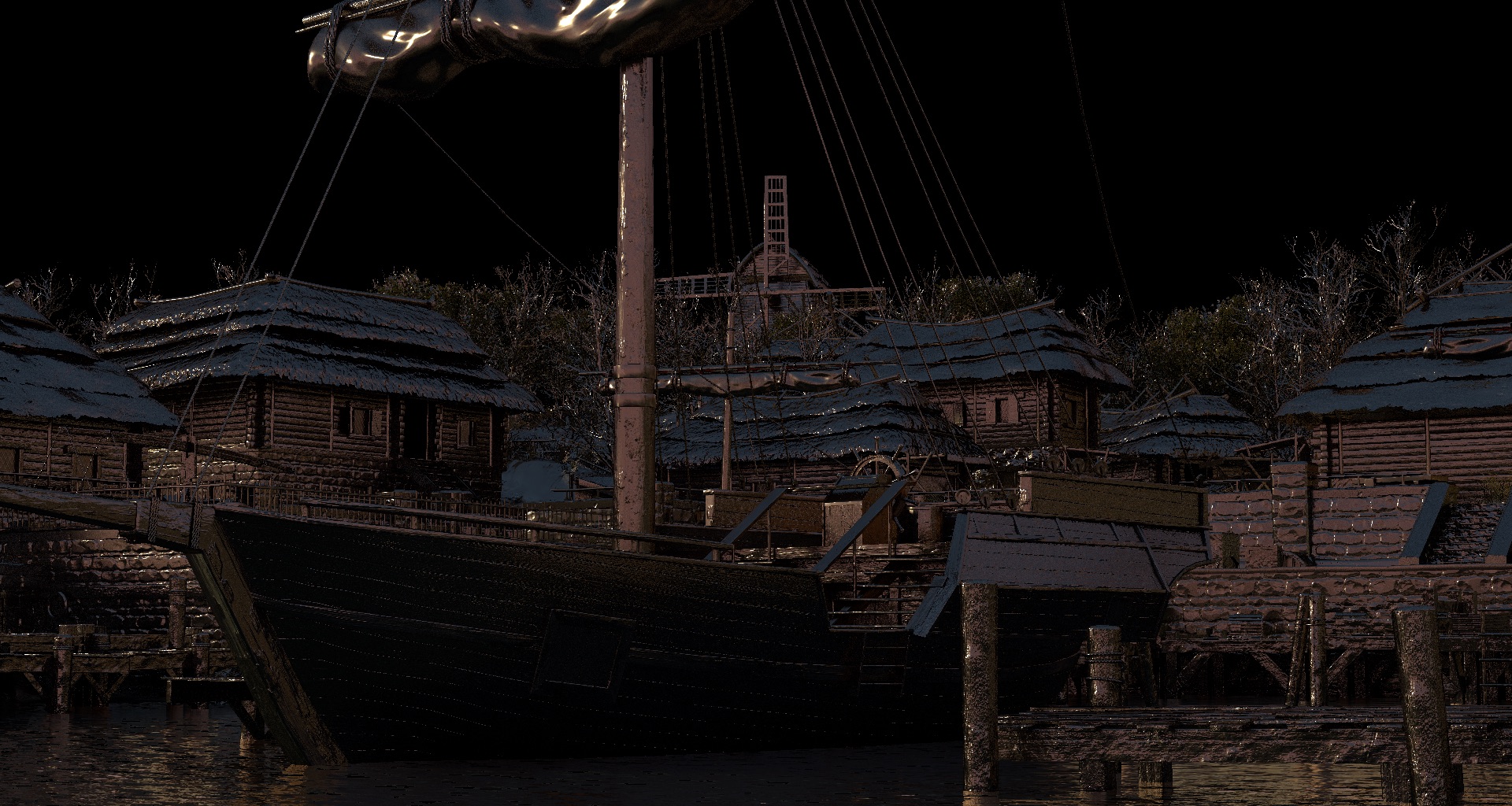}
  \\
  \raisebox{6ex}{$E_3$} & \includegraphics[width=0.31\linewidth]{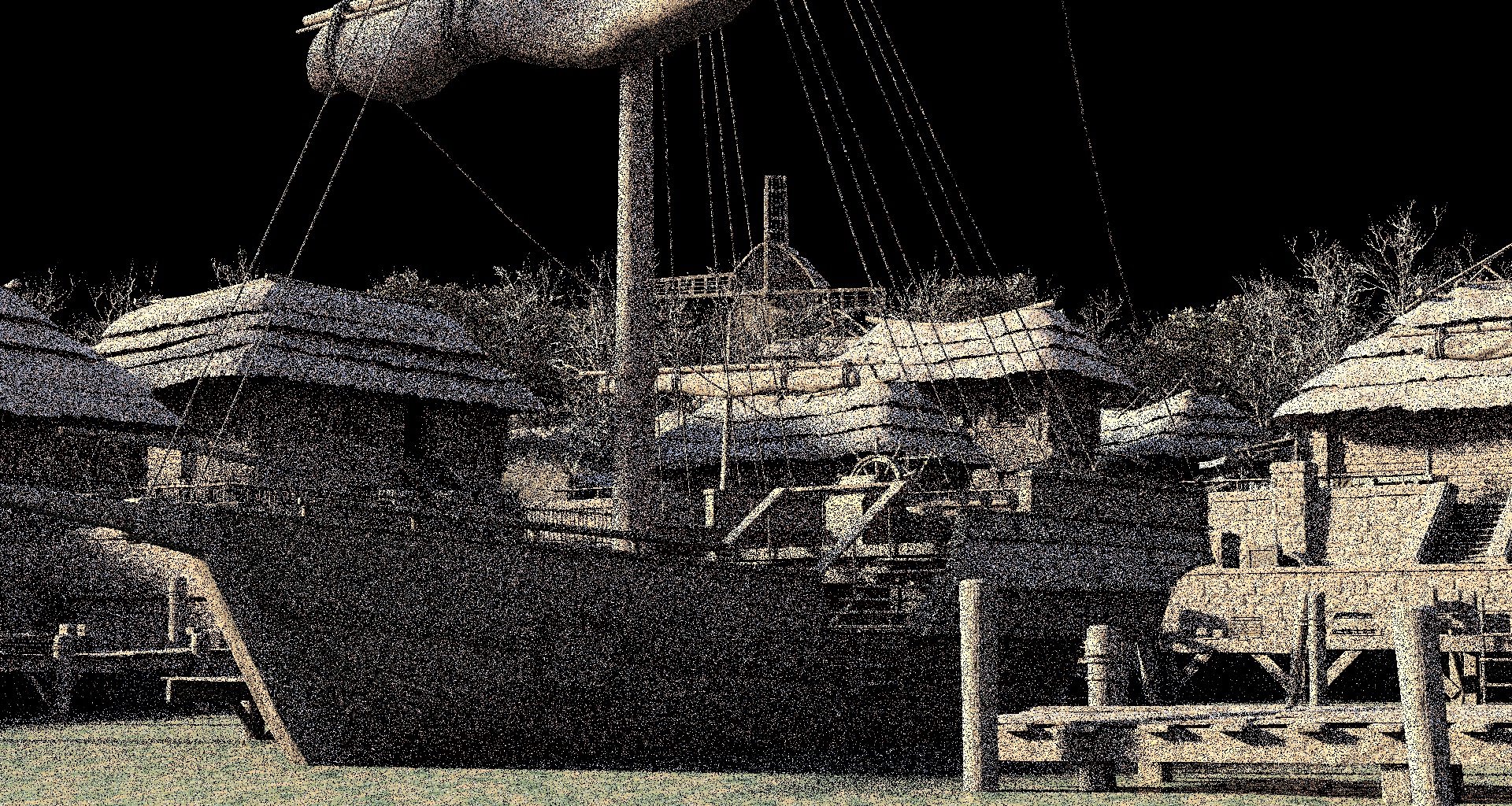}
  & \includegraphics[width=0.31\linewidth]{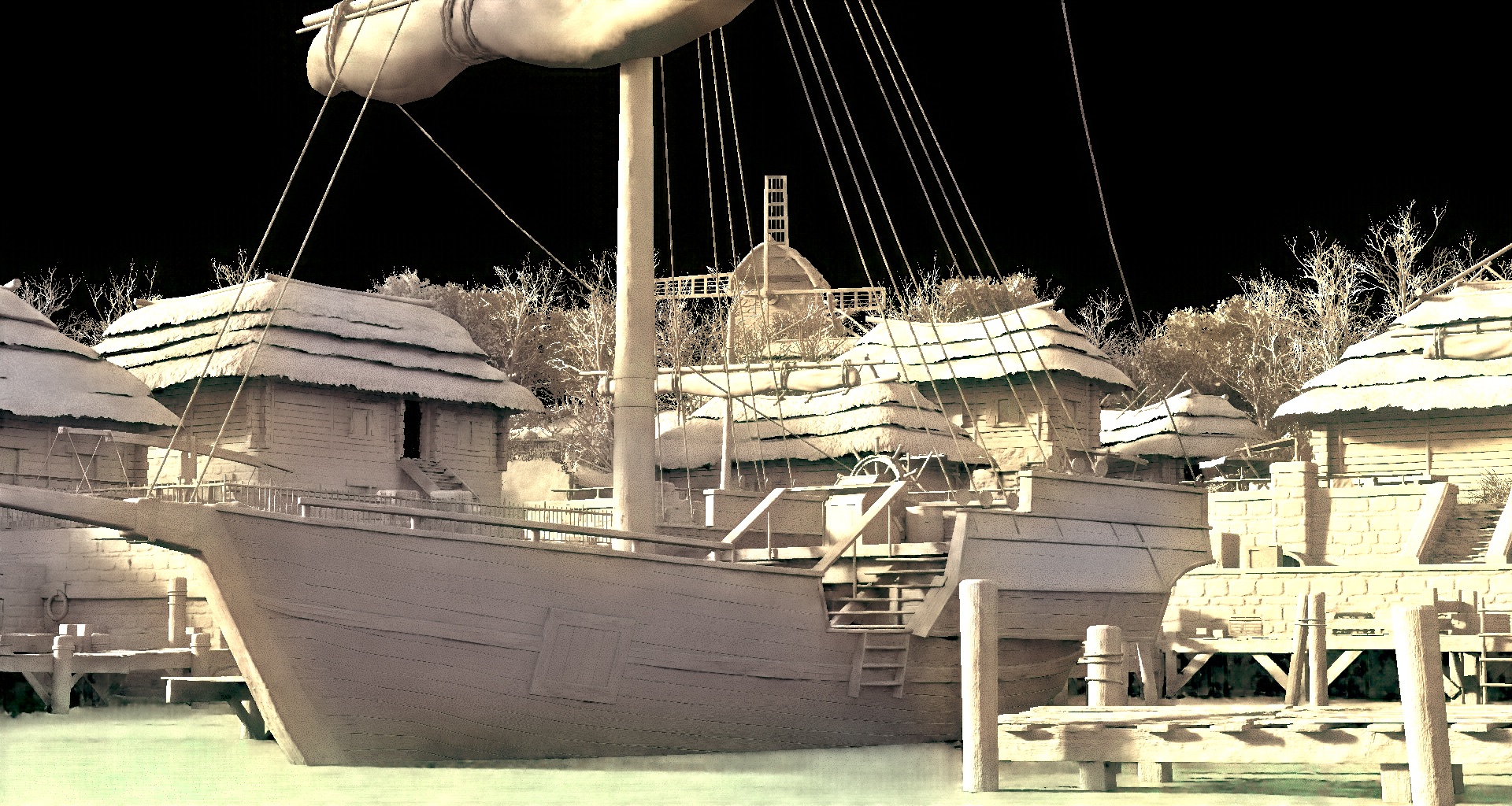}
  & \includegraphics[width=0.31\linewidth]{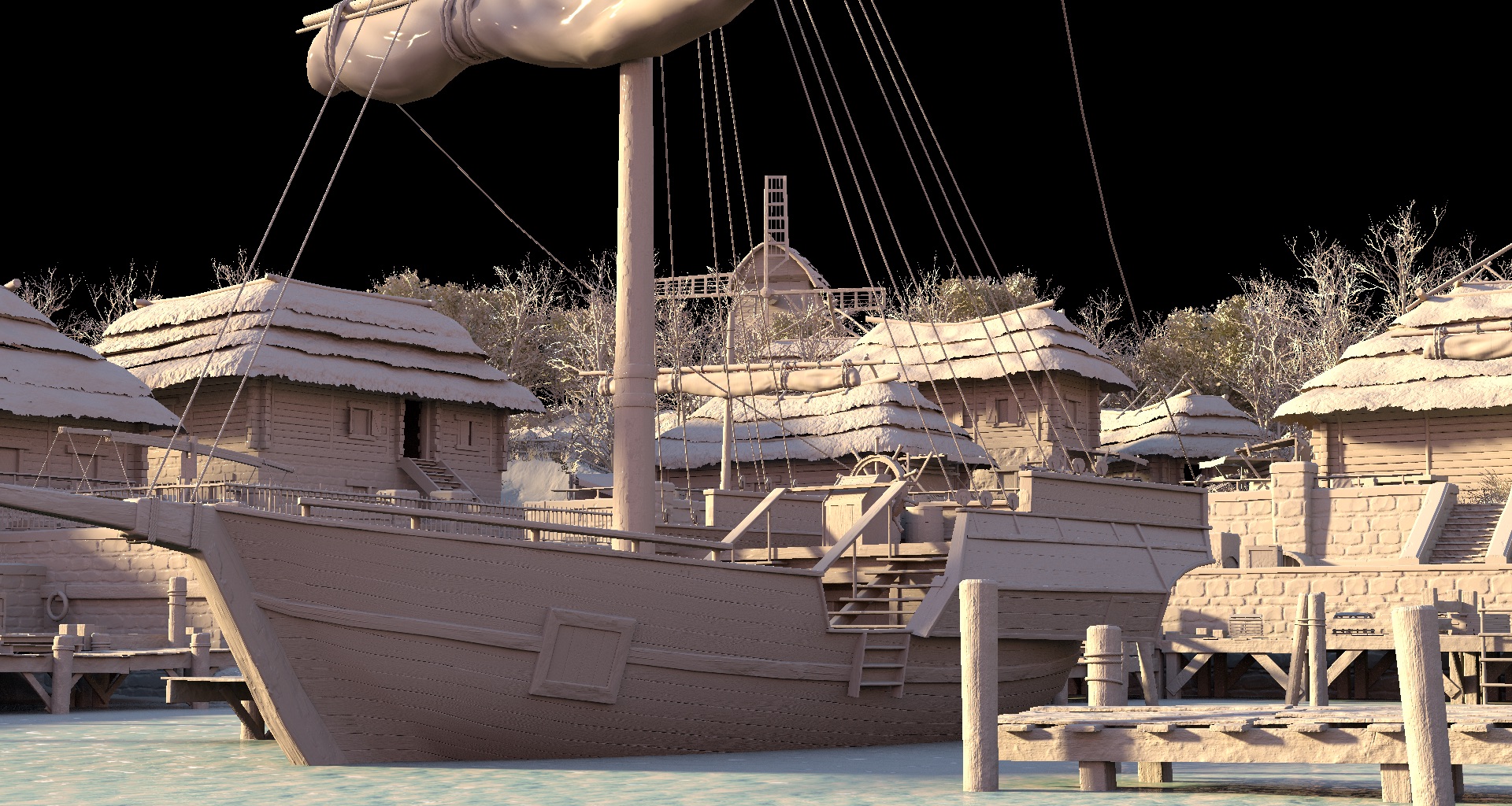}
  \\
  \raisebox{6ex}{$E_4$} & \includegraphics[width=0.31\linewidth]{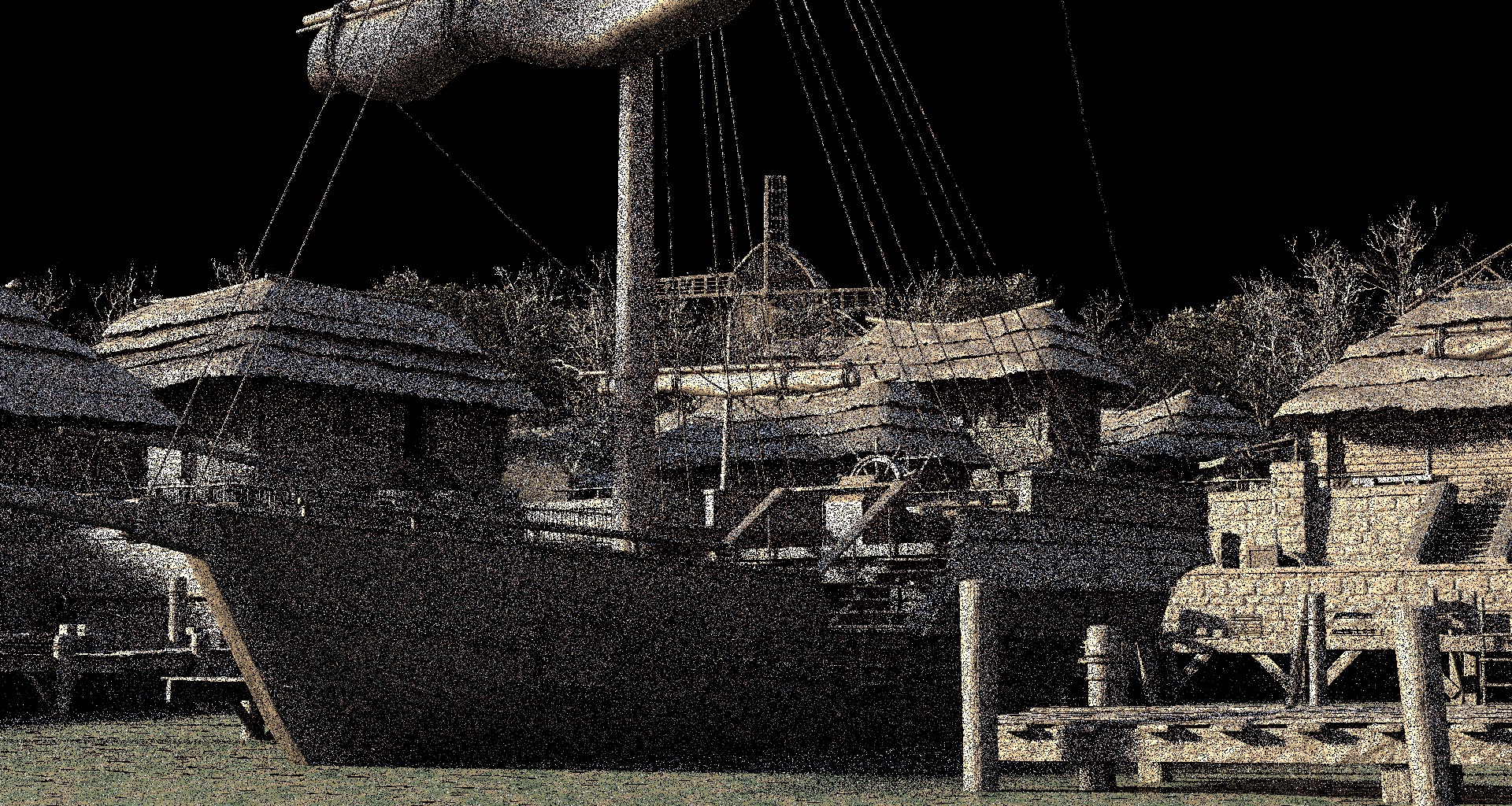}
  & \includegraphics[width=0.31\linewidth]{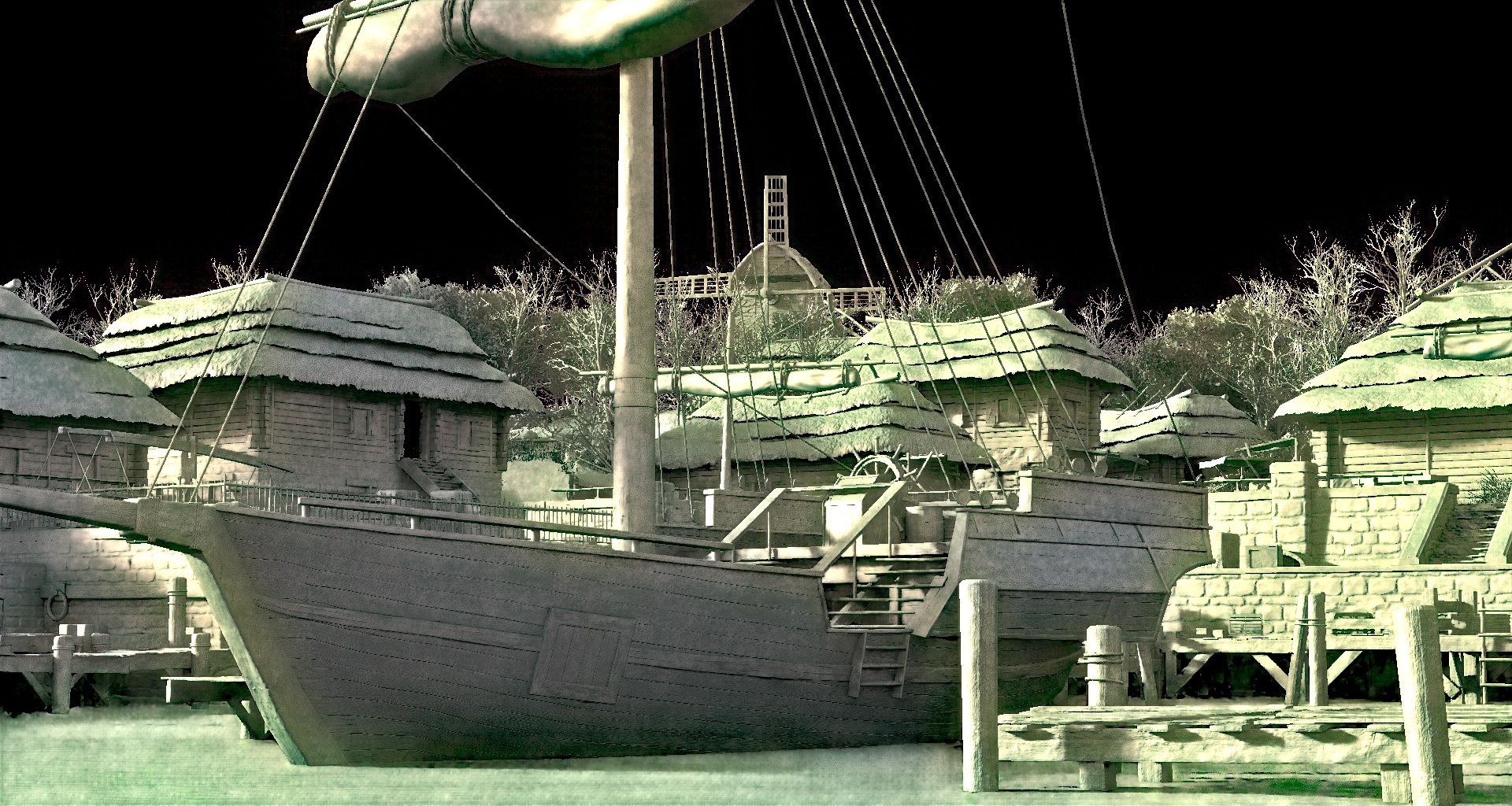}
  & \includegraphics[width=0.31\linewidth]{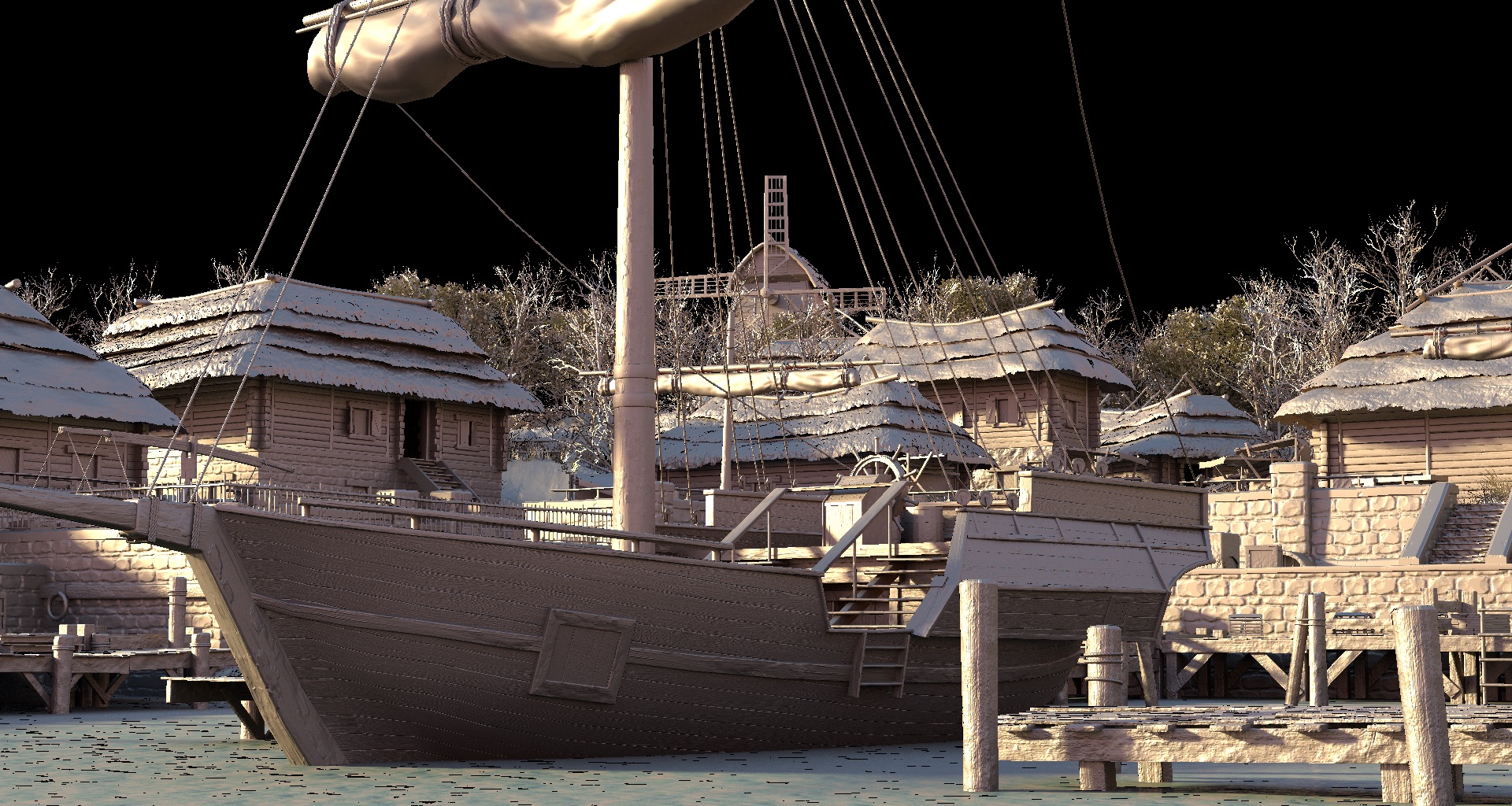}
  \\
  & $\mu$ at 1spp & denoised $\mu'$ & reference $\mu^*$
  \end{tabular}
  \caption{Comparison of the components of the projected irradiance $\mu$ at 1 sample per pixel (1 spp),
  the denoised projected irradiance $\mu'$, and the ground truth reference $\mu^*$.
  Note that the first component (top row) actually is the classic irradiance. The
  geometry is recognizable as the $\cos \vartheta_r$ term and next event estimation
  are included in the
  estimates. Next event estimation contributes shadows from visibility and quadratic
  attenuation with distance.
  The denoiser is trained to take advantage of the geometry clues while filtering.
  Note that the color variations of $\mu'$ as compared to $\mu$ and $\mu^*$ are a
  consequence of how color channel scale is learned by the denoiser and compensated in the material decoder.}
  \label{Fig:ProjectedIrradiance}
\end{figure}

\subsection{Irradiance Projection} \label{Sec:IrradianceEncoder}

The components of the vector-valued function $\bsE(\omega, \omega_r) \in \mathbb{R}^d$
span a $d$-dimensionally, directionally dependent space to represent irradiance. Its dependence on pairs
of directions in the view-aligned tangent space differentiates
$\bsE(\omega, \omega_r)$ from classic vectors of basis functions.
Note that $\bsE$ is independent of spatial location, as neighborhood is implicit across the image plane.

The bidirectional scattering distribution function (BSDF) $f$ in (\ref{Eqn:IEQ}) usually is
mo\-deled as linear combination of a constant function and a mixture of transformed spherical
Gaussians \cite{ShadingFilmGames}. Accordingly, selecting the first component as constant
$E_0(\omega, \omega_r) = 1$ allows for representing averages.
The remaining $d - 1$ components are linearly independent Gaussian lobes created from the Disney
principled BSDF model (Sect.~\ref{Sec:Disney}).
This construction principle for $\bsE(\omega, \omega_r)$ can be used with any other BSDF model, too.
 
The projected irradiance integral in (\ref{Eqn:Approximation}) is evaluated by Monte
Carlo or quasi-Monte Carlo integration~\cite{PBRT}.
Summing the contributions $(R_i, G_i, B_i)$ of the $i$-th path space sample
according to~(\ref{Eqn:IEQ}), we accumulate a concatenated vector
\begin{equation} \label{Eqn:Coefficients}
  \mu := \left[ \frac{1}{n} \sum_{i=0}^{n-1} \bsE(\omega_i, \omega_r) R_i \left| \frac{1}{n} \sum_{i=0}^{n-1} \bsE(\omega_i, \omega_r) G_i  \right| \frac{1}{n} \sum_{i=0}^{n-1} \bsE(\omega_i, \omega_r) B_i  \right]
  \in \mathbb{R}^{3d}
\end{equation}
for each color channel without weighing by the BSDF $f$, also known as
"BSDF stealing"~\cite{ItJustWorks}.
In our notation, the path space sample contributions $(R_j, G_j, B_j)$ are already divided by their respective probability.
Furthermore, all weights of sampling techniques like multiple importance sampling~\cite{PBRT}
for sampling over the hemisphere and next event estimation are included.
As $E_0(\omega, \omega_r) = 1$, the components
\begin{equation} \label{Eqn:Irradiance}
  (\mu_{R, 0},\mu_{G, 0},\mu_{B, 0})
\end{equation}
accumulate the diffuse irradiance~\cite{Irradiance},
which is essential for normalizing purposes later on. 

The irradiance projection is illustrated in Fig.~\ref{Fig:ProjectedIrradiance}.
We sample $n = 1$ path that scatters at the primary hitpoint
as seen from the viewpoint and in the same location sample the light sources for next event
estimation. Both contributions are summed up using their respective multiple importance sampling weights
while omitting the BSDF factor as mentioned before.

\paragraph{Learning from Machine Learning.}

Especially in computer graphics, many functions can be visualized and hence
understood quickly. This provides a new way of discovery:
Understanding what the neural networks are really approximating.
In fact, we explored representing $\bsE$ by 
a ResNet $\bsE_\theta$ with 6 ResNet blocks~\cite{ResNet2} of 16 neurons each.
Visualizations after training revealed that one component remains mostly
constant, while the remaining components approximate the
lobes of a BSDF, very similar to a mixture of spherical Gaussians. 
As a consequence, modeling $\bsE$ analytically as described above (Sect.~\ref{Sec:Disney})
avoids training an additional neural network, resulting in a more efficient and robust algorithm.
Besides, projecting onto functions from the BSDF makes mathematical sense.

\begin{figure}
  \centering
  \begin{tabular}{ccc}
    \includegraphics[width=0.32\linewidth]{1_spp/dim_0}
    & \includegraphics[width=0.32\linewidth]{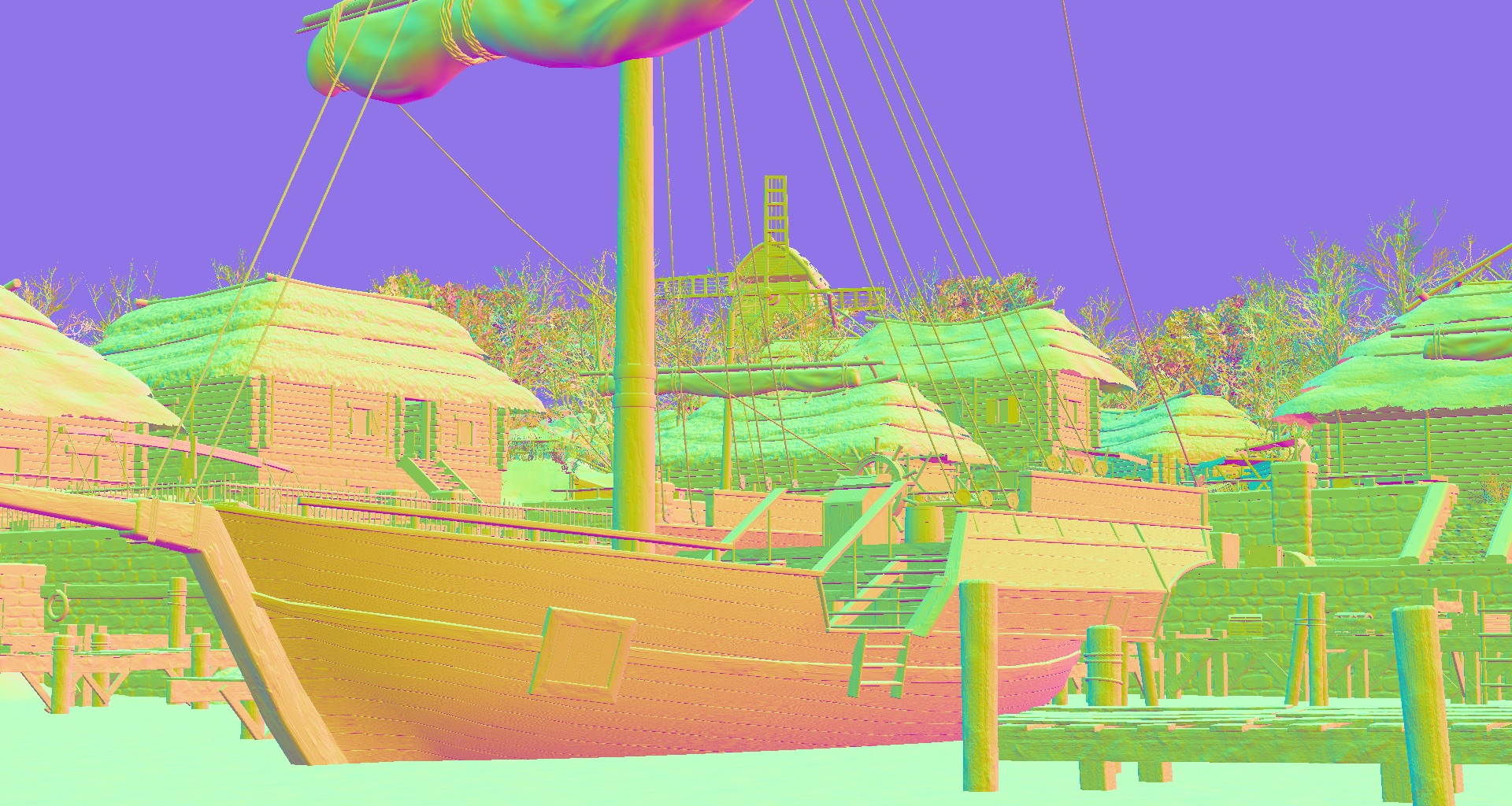}
    & \includegraphics[width=0.32\linewidth]{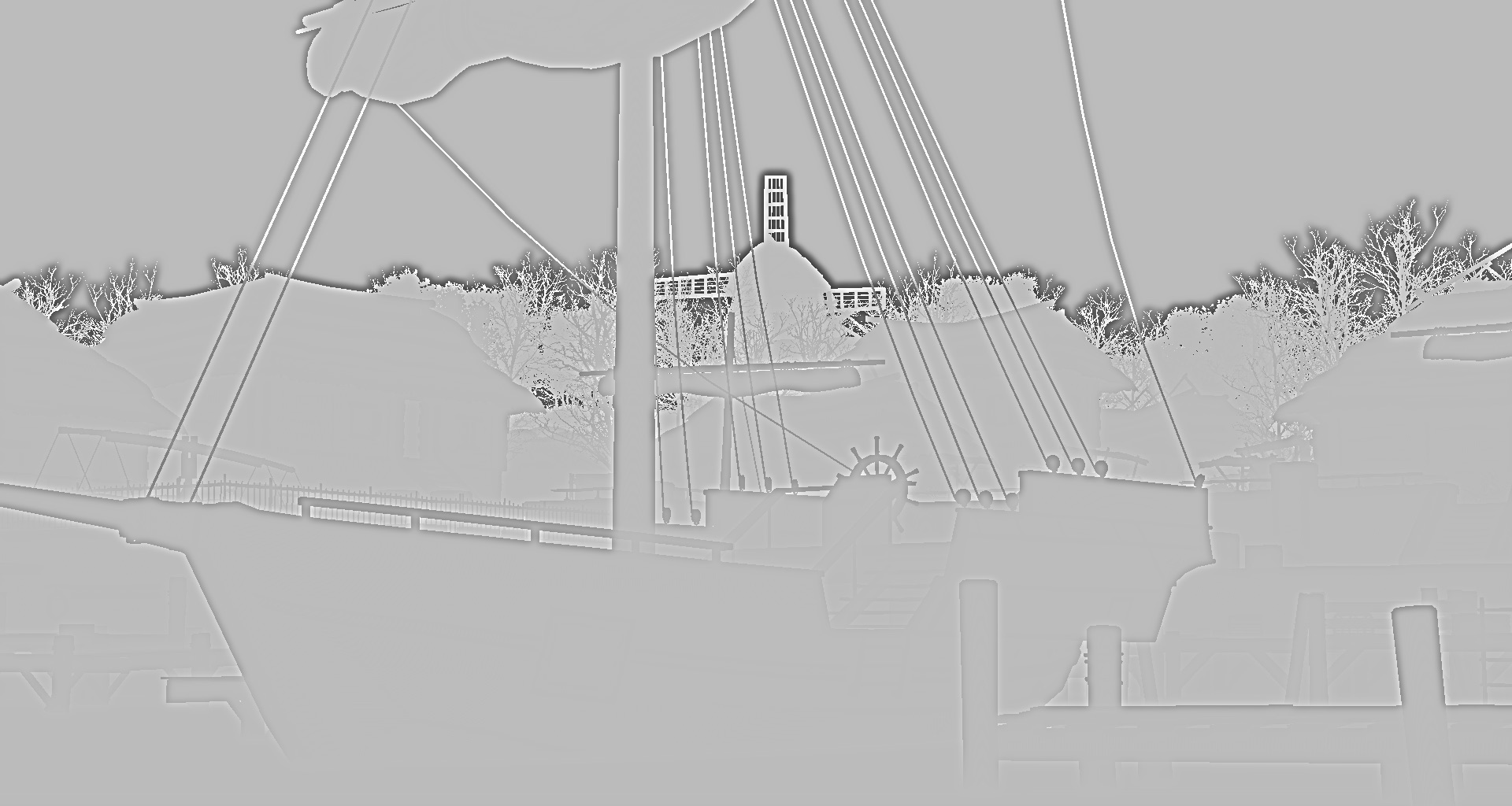} \\
    irradiance & normal $\hat n$ & depth $\hat d$ \\
    $(\mu_{R,0}, \mu_{G,0}, \mu_{B,0})$ \\ \\
    \includegraphics[width=0.32\linewidth]{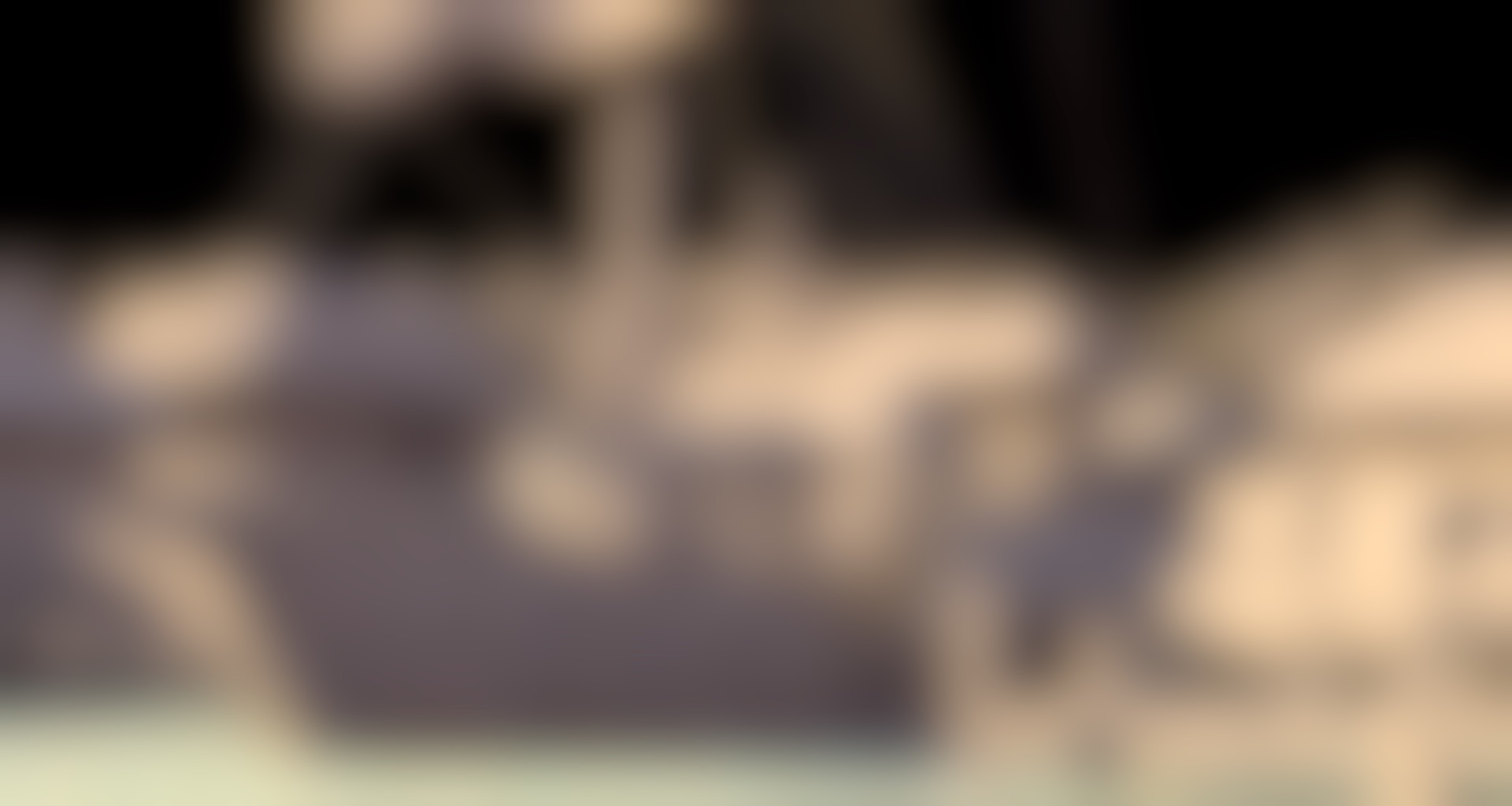}
    & \includegraphics[width=0.32\linewidth]{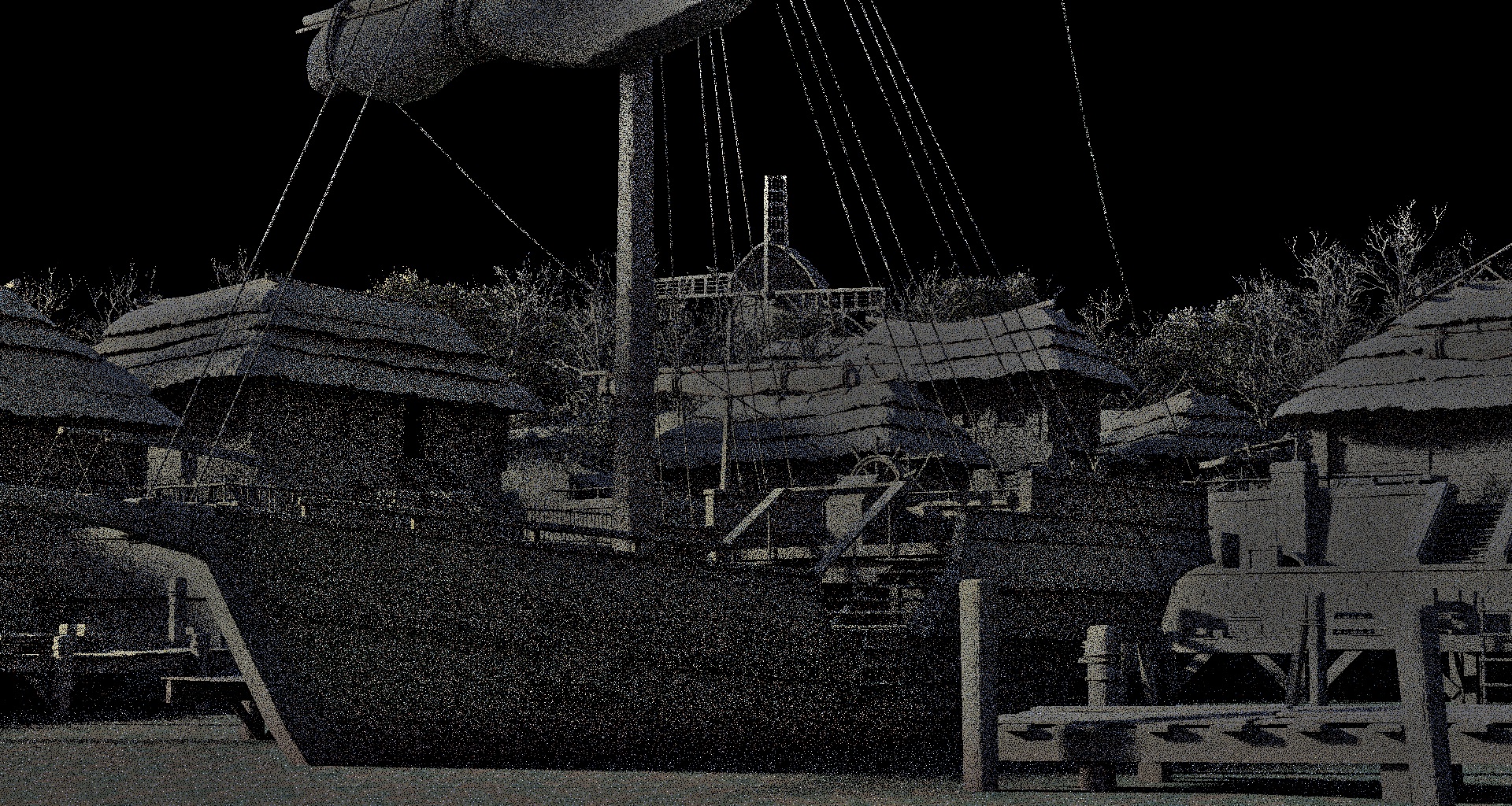}
    & \includegraphics[width=0.32\linewidth]{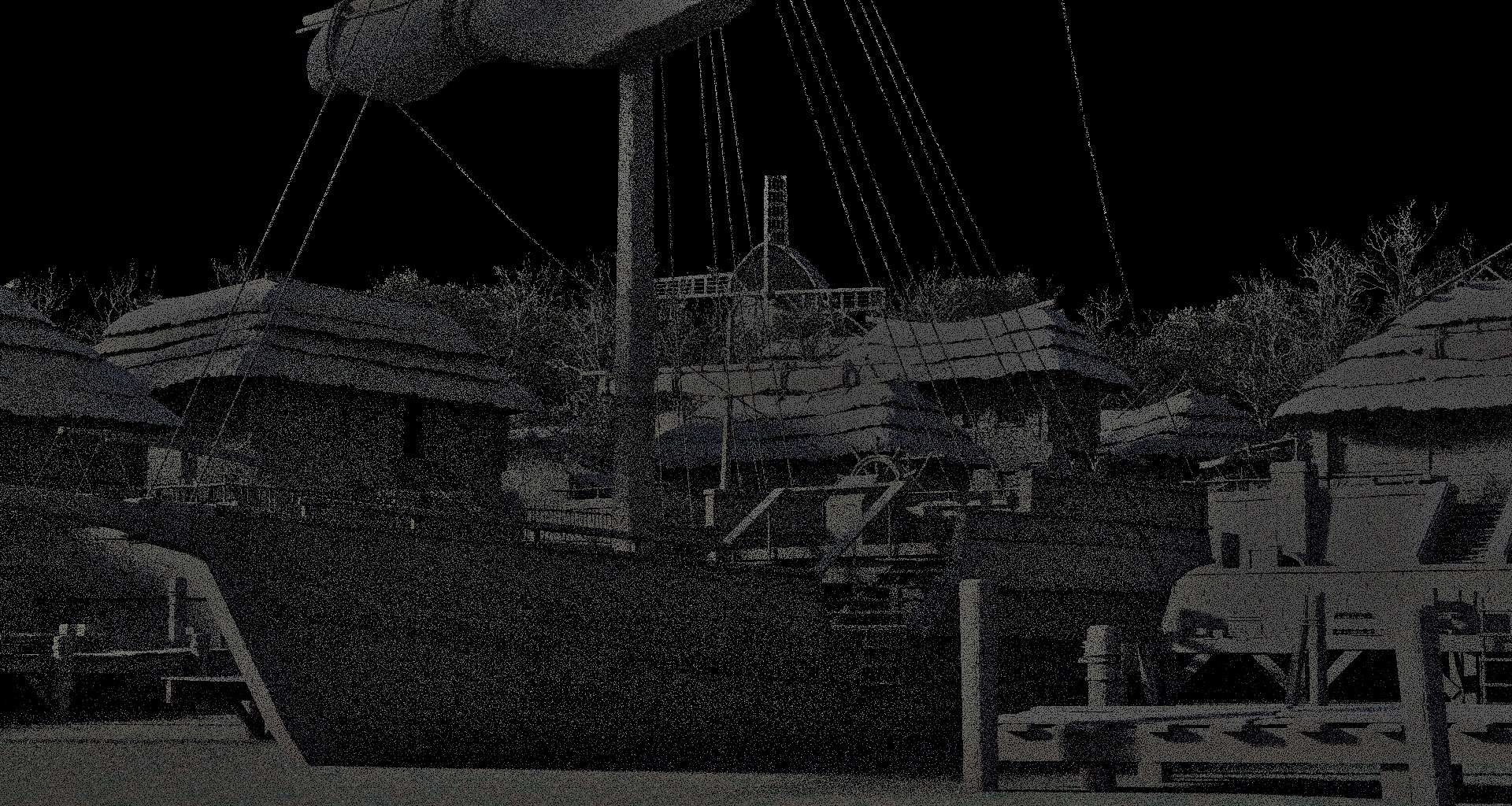} \\
    blurred irradiance & scale removed & compressed range \\
    $(\hat{\mu}_{R,0}, \hat{\mu}_{G,0}, \hat{\mu}_{B,0})$ & $\frac{\mu_{R,0}}{\max\{\hat{\mu}_{R,0}, \epsilon\}}$ & $\log\left(1 + \frac{\mu_{R,0}}{\max\{\hat{\mu}_{R,0}, \epsilon\}}   \right)$
  \end{tabular}
  \caption{Visualization of the denoiser inputs irradiance $(\mu_{R,0}, \mu_{G,0}, \mu_{G,0})$, normals, depth, and the 
  the blurred irradiance $(\hat \mu_{R,0}, \hat \mu_{G,0}, \hat \mu_{G,0})$
  that is practically noise-free.
  Dividing the irradiance by the blurred irradiance normalizes scale, yielding almost gray images.
  The remaining high dynamic range noise is compressed by applying
  a logarithm. Note that black components remain black after the transformation.}
  \label{Fig:DenoiserInputs}
\end{figure}

\begin{figure}
\centering
\includegraphics[width=\linewidth]{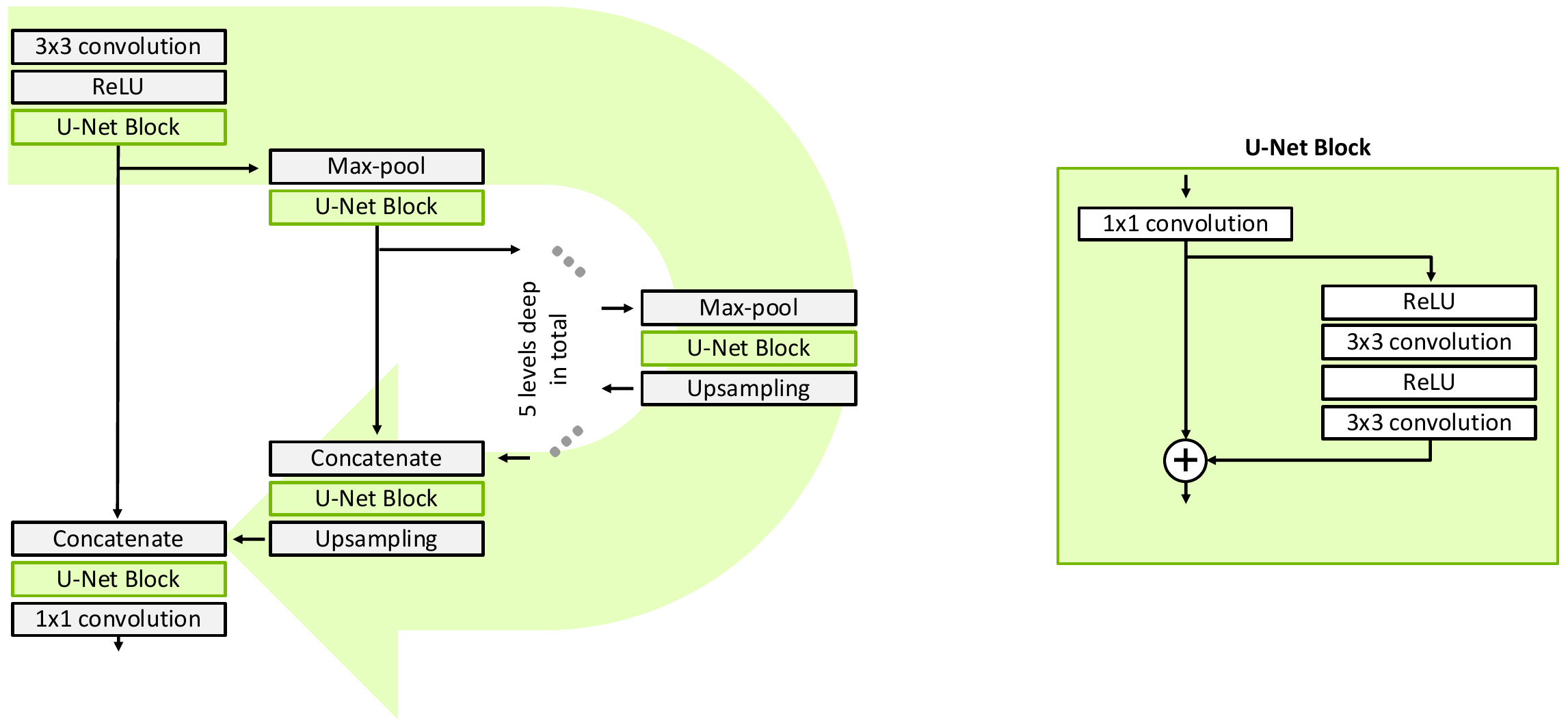}
\caption{As compared to the original U-Net \cite{U-Net}, we only use one convolution
per layer and ResNets~\cite{ResNet2} for the stacked 1x1 convolutions, resulting in a more efficient architecture.
The network has about 9 million weights.}
\label{Fig:U-Net}
\end{figure}

\subsection{Denoising the Projected Irradiance} \label{Sec:Denoiser}

In order to overcome blur due to filtering shaded images,
we filter noise in projected irradiance space before shading. Then,
shading after filtering (Sect.~\ref{Sec:MaterialDecoder})
adds texture detail. This approach results in crisper images,
because detail is not masked by noise.
In principle, the variance of the parametric projected irradiance $\mu$ in
(\ref{Eqn:Coefficients}) is reduced by applying a U-Net $U_\theta$ \cite{U-Net} (see Fig.~\ref{Fig:U-Net})
with the objective (Sect.~\ref{Sec:Training}) to make it a better estimate of the projected irradiance
integral in (\ref{Eqn:Approximation}).

Previous denoising work in graphics \cite{Bako17,Noise2Noise} teaches that neural networks
do not cope well with high dynamic range input like the projected irradiance estimate $\mu$.
We address this issue by two subsequent, invertible transformations: First, we remove the scale from the
irradiance by dividing it by a local average of the irradiance. While then free of scale,
the remaining noise still may have a high dynamic range and hence is compressed
using a logarithm \cite[Sect.~4.3]{Bako17}.
Note that opposite to division by the average, subtracting the average would not remove the scale.

Ideally, we would normalize $\mu$ using the true irradiance from $\mu^*$,
which would perfectly remove scale, but obviously is not available. However,
scaling $\mu$ with the reciprocal local estimate of the average irradiance
is feasible. Since the same estimate must be used to restore brightness after
the non-linear denoising operation, it needs to be practically noise-free. 
Applying a wide Gaussian filter approximated by \`a-trous wavelets \cite{FastAtrous}
to the sampled irradiance yields the desired local averages to remove the scale.
To efficiently compute the noise-free blurred irradiances
$(\hat{\mu}_{R, 0}, \hat{\mu}_{G, 0}, \hat{\mu}_{B, 0})$,
the irradiances $(\mu_{R, 0}, \mu_{G, 0}, \mu_{B, 0})$
are blurred by simply iterating an \`a-trous filter \cite{FastAtrous} with
a $5 \times 5$ Gaussian filter kernel
6 times to rapidly achieve a sufficiently large blur radius.
We do not apply common edge stopping heuristics \cite{ATrous},
as otherwise fine geometry may become masked by noise.

The implementation starts by normalizing the projected irradiance vectors
by their respective accumulated irradiances (\ref{Eqn:Irradiance}).
Concatenated across the color channels, we have
\begin{eqnarray}
  \lefteqn{\text{normalize}( \mu)}  \nonumber \\
  & := & \left[ \frac{(\mu_{R,0}, \ldots, \mu_{R,d - 1})}{\max\{\mu_{R,0}, \epsilon\}} \left| \frac{(\mu_{G,0}, \ldots, \mu_{G,d - 1})}{\max\{\mu_{G,0}, \epsilon\}} \right| \frac{(\mu_{B,0}, \ldots, \mu_{B,d - 1})}{\max\{\mu_{B,0}, \epsilon\}} \right] , \label{Eqn:Normalization}
\end{eqnarray}
where a small $\epsilon$ avoids numerical issues in very dark regions,
especially divisions by zero. Note that whenever a component of $(\mu_{R, 0},\mu_{G, 0},\mu_{B, 0})$
is larger or equal to $\epsilon$, its normalization is one, and
smaller otherwise. Zero components, which are quite common among colors, remain zero.
The normalization removes the brightness and hence color from the components, leaving vectors of
ratios of the components. This is necessary, as the color and hence brightness later on is restored
using the blurred irradiance $\hat \mu$.
Exemplary for the red channel, the denoiser input then is
\begin{equation} \label{Eqn:Compress}
  \mu^c_R = \frac{(\mu_{R,0}, \ldots, \mu_{R,d - 1})}{\max\{\mu_{R,0}, \epsilon\}}  \cdot \log\left(1 + \frac{\mu_{R,0}}{\max\{\hat{\mu}_{R,0}, \epsilon\}}   \right)
\end{equation}
and Fig.~\ref{Fig:DenoiserInputs} visualizes the input transformation that consists of
scale removal and compression. Note that the logarithmic factor is zero whenever the irradiance
is zero.

To help the U-Net $U_\theta$ \cite{U-Net}  (see Fig.~\ref{Fig:U-Net}) gather information across pixels, we add geometric cues.
The normal $\hat n$ in $x$ is transformed into view space in which it becomes
independent of  camera pose, allowing the U-Net to consider changing
normals, for example, as present around edges.
Information beyond the one provided by normals, may be revealed in the depth buffer.
The range of depth values is highly scene dependent and a high dynamic range may not work
nicely with a trained denoiser. Hence, instead of
mapping the values by scene-dependent scales or using minimum and maximum
values of the depth buffer, the input to the neural network is prepared in a way
similar to the irradiance compression:
The distance from the view point to $x$ is blurred by
iterating the aforementioned \`a-trous filter for 3 times.
Each distance then is divided by its corresponding blurred depth,
centered at zero by subtracting one, and clamped to the range of [-1,1]
yielding the range compressed depth buffer $\hat d$.
Hence, for a constant, non-zero depth, where the depth equals the blurred depth, the range compressed depth buffer
is zero. Otherwise, the sign indicates the direction of gradients in depth to the network.

The denoiser as a convolutional neural network locally fixes brightness
and hierarchically computes weighted averages across neighbors. 
Its output 
\begin{equation} \label{Eqn:Denoiser}
  \mu' = (\hat{\mu}_{R,0}, \hat{\mu}_{G,0}, \hat{\mu}_{B,0}) \odot U_\theta\left(\mu^c, \hat n, \hat d\right)
\end{equation}
is the output of the U-Net $U_\theta$ scaled by the blurred irradiance to restore the dynamic range and color.
Here $\odot$ denotes the multiplication of each color component (similar to (\ref{Eqn:Normalization}))
by its respective blurred irradiance. Note that the blurred irradiance is practically noise-free,
while a multiplication with the sampled irradiances (\ref{Eqn:Coefficients}) would add back the noise.
Similar to Bako et al. \cite{Bako17}, the neural network is trained to invert the logarithmic
term in (\ref{Eqn:Compress}).
Reconstructing the image brightness is possible, because
the \`a-trous transform preserves the average image brightness (the integral).

\paragraph{Intuition:}
To understand the model, let us assume that $\mu^* = \mu = \mu'$, meaning
that the projected irradiance is already denoised and hence
\[
\mu_R
= \hat{\mu}_{R,0} \cdot U_\theta \left( \frac{\mu_R}{\max\{\mu_{R,0}, \epsilon\}}  \cdot \log\left(1 + \frac{\mu_{R,0}}{\max\{\hat{\mu}_{R,0}, \epsilon\}}   \right), \hat n, \hat d \right)
.
\]
Looking at the first component $\mu_{R, 0}$,
for $\epsilon \rightarrow 0$ in the limit we then have
\[
\mu_{R, 0}
= \hat{\mu}_{R,0} \cdot U_\theta \left( \log\left(1 + \frac{\mu_{R,0}}{\hat{\mu}_{R,0}}   \right), \hat n, \hat d \right) ,
\]
meaning that for $\mu_{R, 0} >0$ the denoiser $U_\theta$ learns to invert the logarithm and the addition, while
the multiplication by the average $\hat{\mu}_{R,0}$ is explicitly modeled. For the case $\hat{\mu}_{R,0} > 0$
of a non-zero average and $\mu_{R, 0} = 0$, the logarithm returns zero resulting in a zero
output for a scale-invariant U-Net. Finally, if $\hat{\mu}_{R,0} = 0$, $\mu_{R, 0}$ must be zero, too,
because otherwise the average cannot be zero.
In that case we programmatically define that the factor $\hat{\mu}_{R,0} = 0$ zeroes out the result
of the U-Net. Note that using a function like $\tanh$ or a tone mapping operator like
$\frac{x}{1 + x}$ \cite{Reinhard:2002:PTR} instead of the logarithm results in worse
performance, because these functions reach their asymptotes  so
quickly that their inversion becomes a numerical issue.

\subsection{Material Decoder} \label{Sec:MaterialDecoder}

Now that the incident radiance is practically noise-free in projected irradiance space,
the shading is computed. More precisely,
the material decoder $M_\theta$ acts as a neural integral operator
on the projected irradiance (\ref{Eqn:Denoiser})
and approximates $L_r(x, \omega_r)$ by (\ref{Eqn:Approximation})
according to the material evaluated in $x$ and viewing direction $\omega_r$.

\begin{figure}
  \centering
  \begin{tabular}{cccccc}
  &&& black & base color & white \\
  \includegraphics[width=0.157\linewidth]{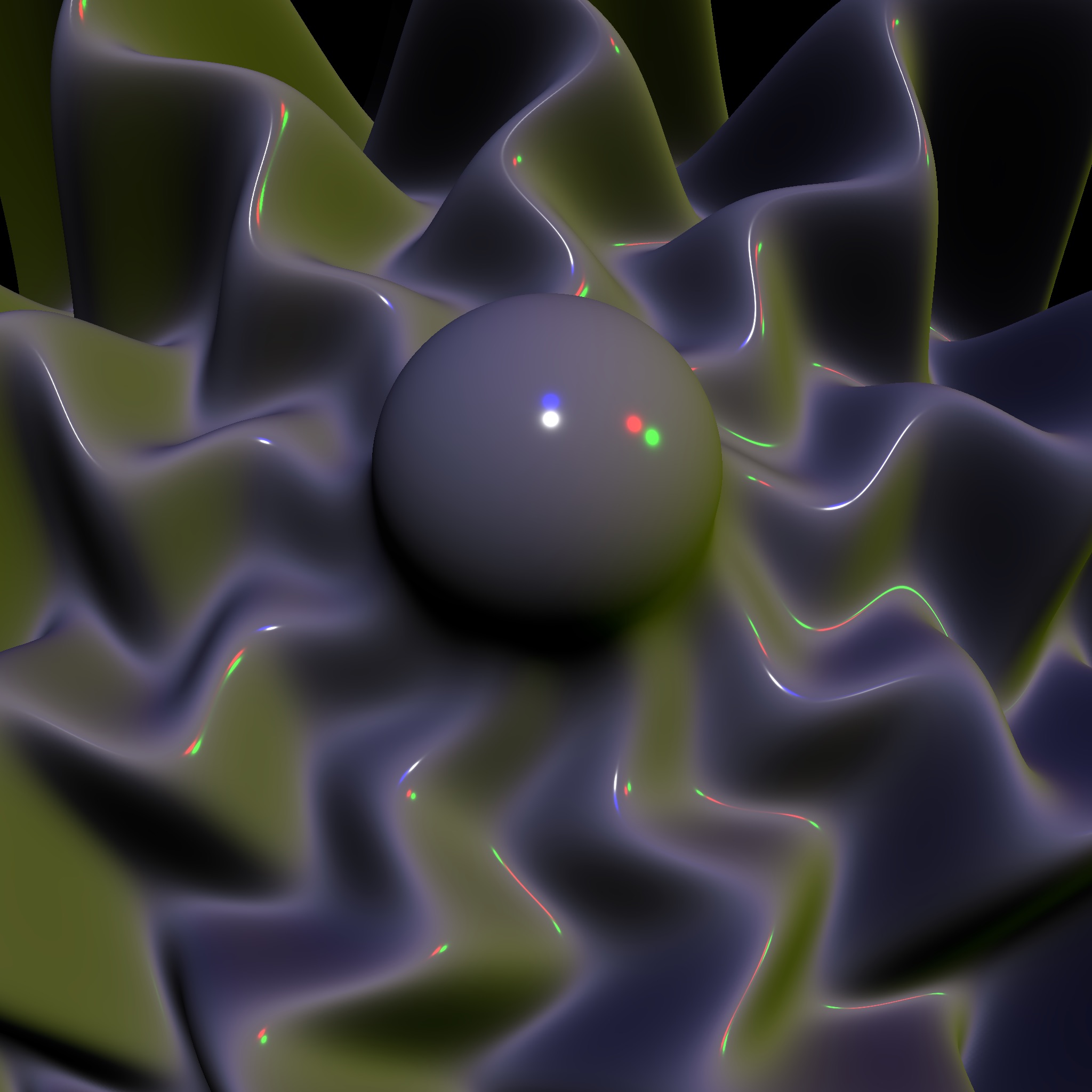}
  & \includegraphics[width=0.157\linewidth]{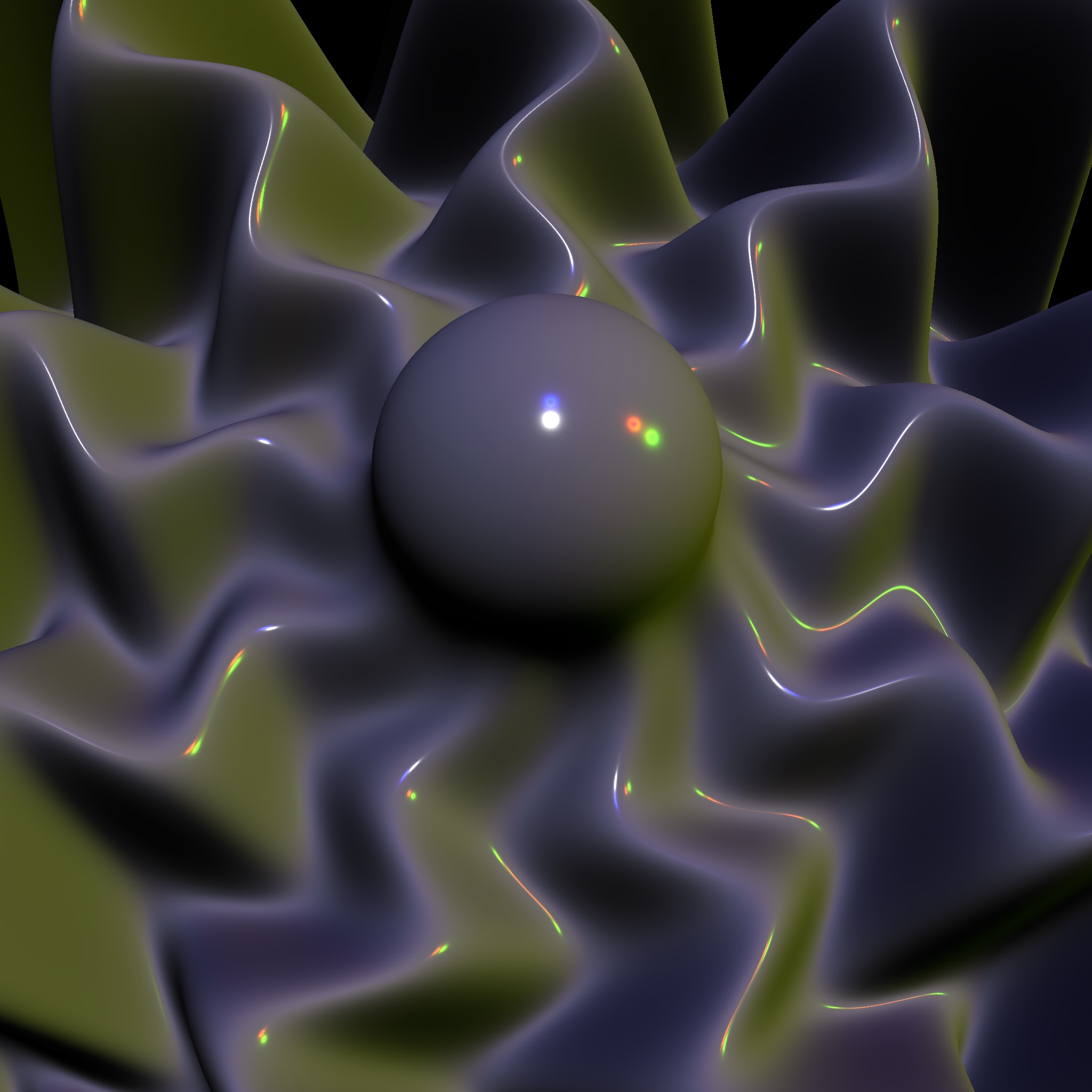}
  & \includegraphics[width=0.157\linewidth]{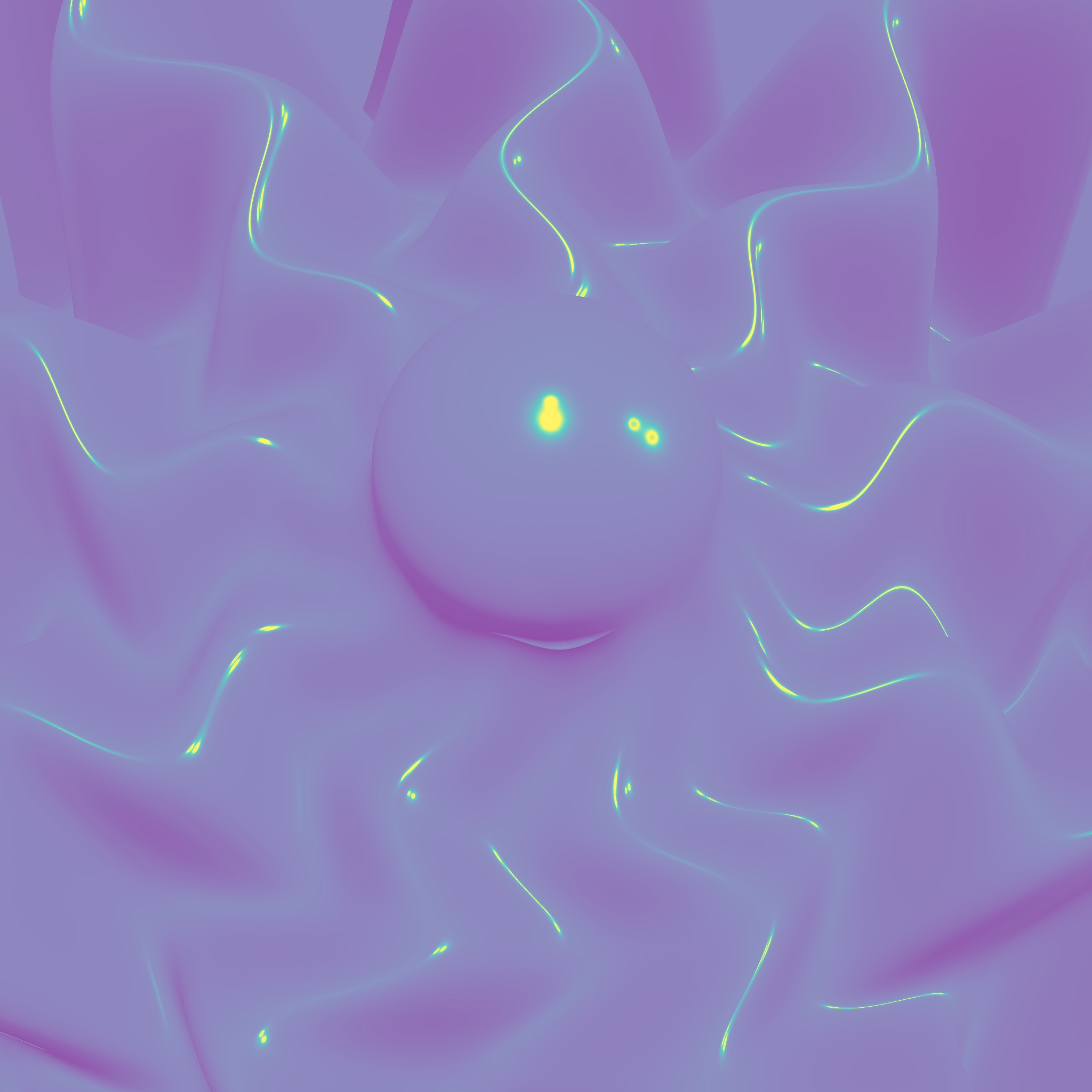}
  & \includegraphics[width=0.157\linewidth]{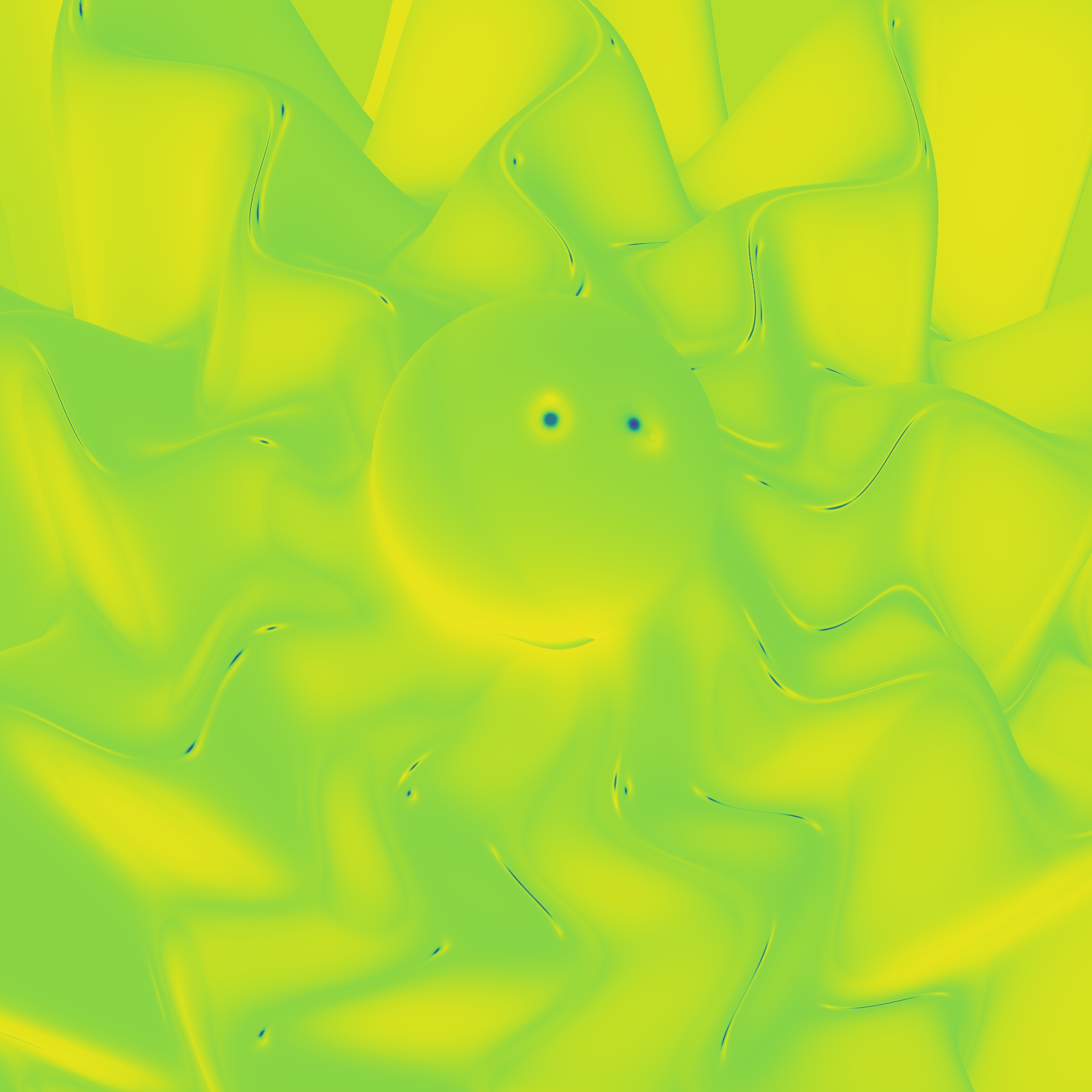}
  & \includegraphics[width=0.157\linewidth]{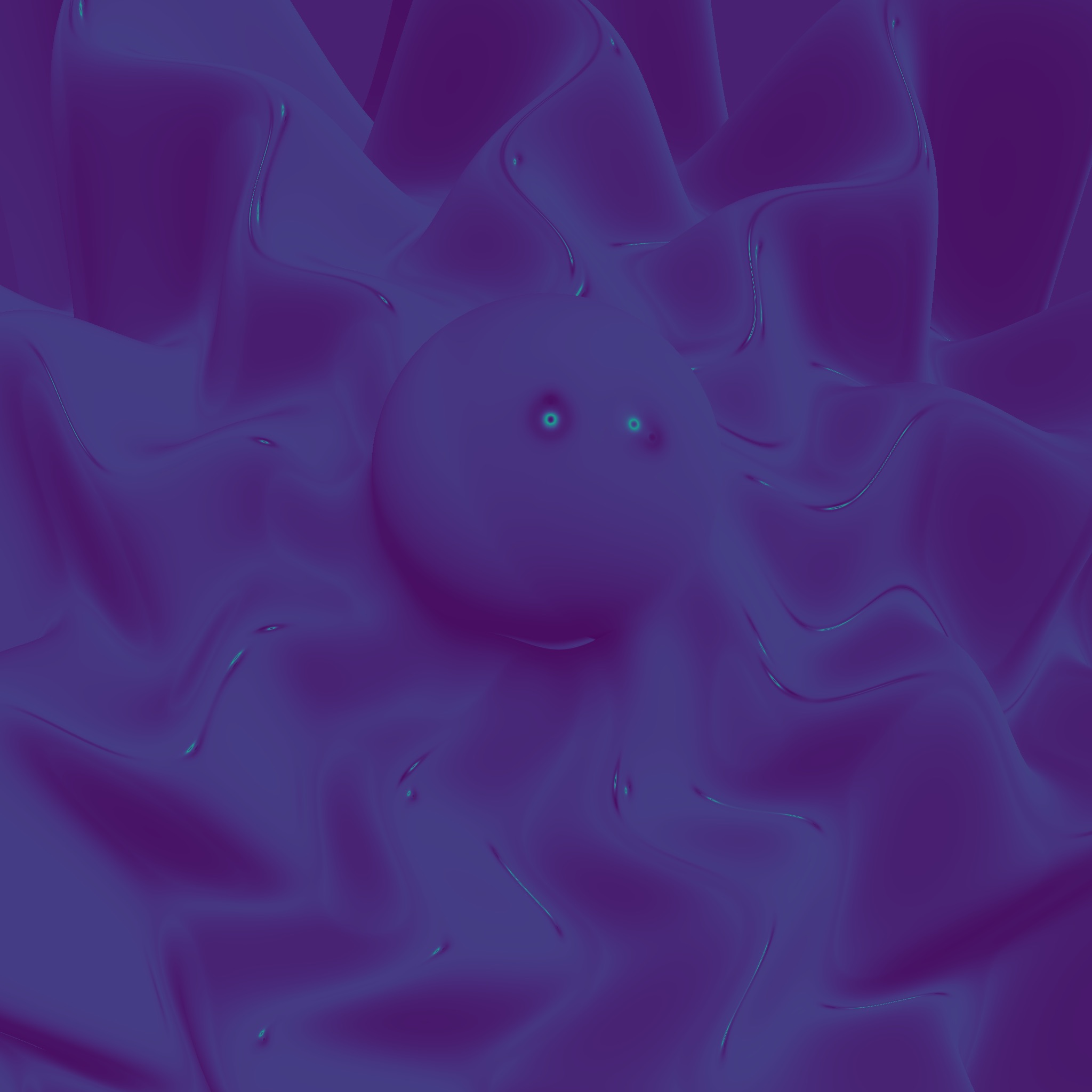}
  & \includegraphics[width=0.157\linewidth]{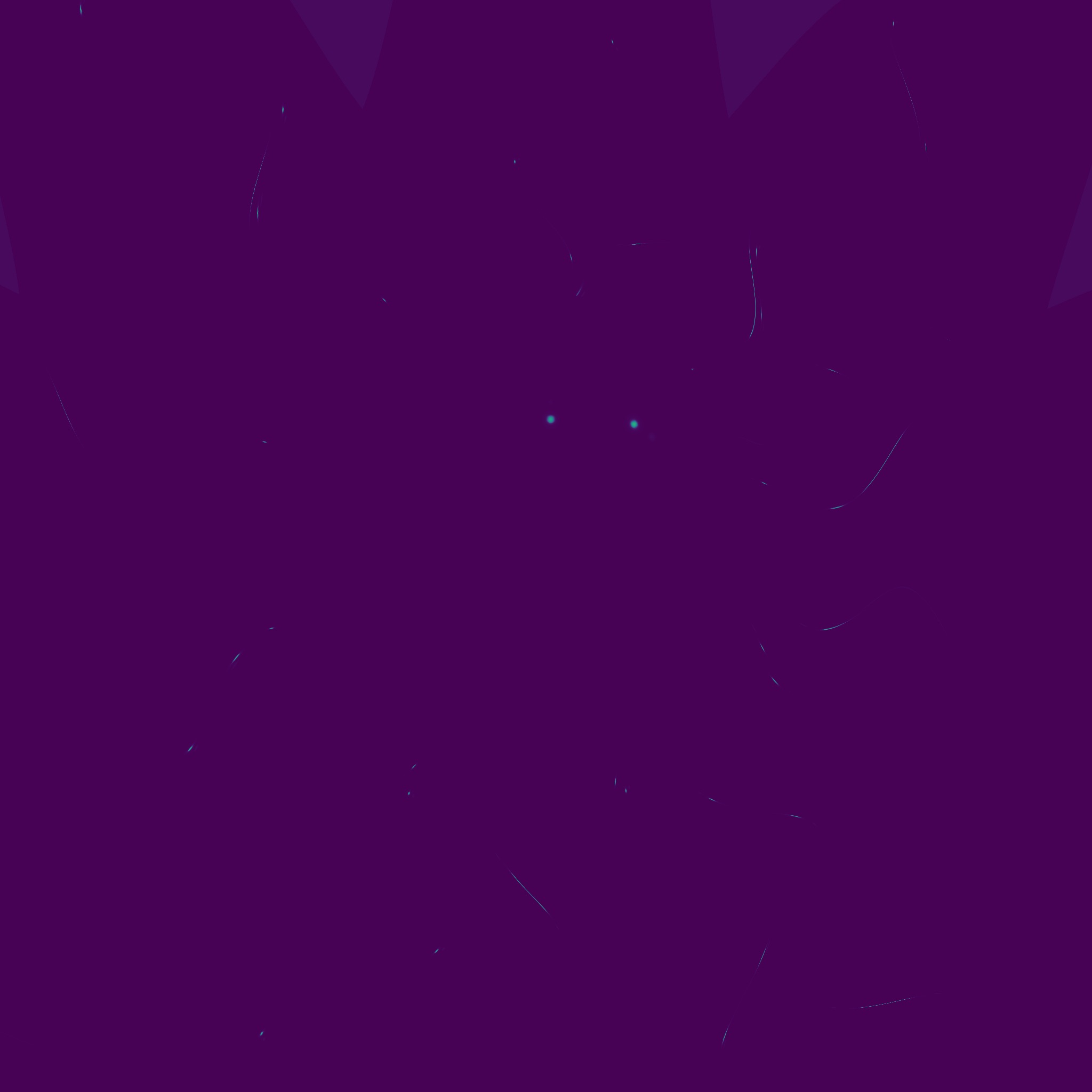} \\
  reference & material  & $I$ & $\hat w_{R,0}$ & $\hat w_{R,1}$ & $\hat w_{R,2}$ \\
  & decoder
  \end{tabular}
  \caption{The image produced by the material decoder $M_{\theta, R}$ in (\ref{Eqn:Interpolation}) is quite close to the reference.
  For illustration, we show a
  false color visualization of the normalized outputs
  and the intensity $I$ of the material decoder neural network (\ref{Eqn:Weights})
  as used in the convex combination (\ref{Eqn:Interpolation}) (exemplary for the red channel $R$ only).}
  \label{Fig:ConvexCombination}
\end{figure}

For the same reasons as before in (\ref{Eqn:Normalization}), the denoised
projected space components are normalized
\begin{equation} \label{Eqn:Normalization2}
  \mu'' = \text{normalize}( \mu') := \left[ \frac{\mu'_{R,1\ldots d-1}}{\max\{\mu'_{R,0}, \epsilon\}} \left| \frac{\mu'_{G,1\ldots d-1}}{\max\{\mu'_{G,0}, \epsilon\}} \right| \frac{\mu'_{B,1\ldots d-1}}{\max\{\mu'_{B,0}, \epsilon\}} \right]
\end{equation}
in order to decouple them from brightness, because it increases the
efficiency of neural network processing.
Given the shading parameters $P(x, \omega_r)$, the neural network
\begin{equation} \label{Eqn:MaterialNetwork}
  m_\theta(\mu'', P(x, \omega_r)) \rightarrow (\hat w_R, \hat w_G, \hat w_B, I) \in \mathbb{R}^{3 \times 3 + 1} ,
\end{equation}
computes three sets of three weights and an intensity $I$. The shading parameters $P(x, \omega)$ consist
of the material parameters and the $\cos \vartheta_r$ specified by the angle $\vartheta_r$ between the viewing direction $\omega_r$
and the surface normal $\hat n$, all evaluated in $x$.
While the material base color $\rho$ is normalized by its $L^2$-norm, the other material parameters
like roughness, specularity, and metallicity are mapped to the range $[-1,1]$ from their respective ranges
for improved neural network performance.
Then, each set of three weights is normalized by a softmax function
\begin{equation} \label{Eqn:Weights}
  (w_R, w_G, w_B) = (\text{softmax}(\hat w_R), \text{softmax}(\hat w_G), \text{softmax}(\hat w_B))
\end{equation}
such that weights per color channel sum up to one.
Exemplary for the red component, the material decoder
\begin{eqnarray}
  M_{\theta, R}(x, \omega_r, \mu'_R) & := & \mu'_{R, 0} \cdot I \cdot (\overbrace{0 \cdot w_{R,0}}^{\text{black}} + \overbrace{w_{R,1} \cdot \rho_R}^{\text{base color}} + \overbrace{1 \cdot w_{R,2}}^{\text{white}}) \nonumber \\
    & = & \mu'_{R, 0} \cdot I \cdot (\underbrace{w_{R,1} \cdot \rho_R + w_{R,2}}_{\in [0,1]}) . \label{Eqn:Interpolation}
\end{eqnarray}
is modeled as a convex combination of black ($= 0$),
the material base color $\rho \equiv \rho(x) \in [0,1]^3$ (albedo, this time not normalized) evaluated in the shading point $x$, and white ($= 1$)
that is scaled by an intensity $I$ and the filtered irradiance $\mu'_{R, 0}$ from (\ref{Eqn:Denoiser})
as depicted in Fig.~\ref{Fig:ConvexCombination}.
The convex combination allows for representing any color in the RGB unit cube,
where black and white can be mixed into the base color of the material.
The intensity $I$ is shared across color channels,
while the filtered irradiance restores the range per color channel.
This explains why the components $(\mu'_{R, 0},\mu'_{G, 0},\mu'_{B, 0})$ were
excluded from normalization in (\ref{Eqn:Normalization2}).
Modeling the products explicitly avoids teaching multiplication to neural networks.

The neural network $m_\theta$ in (\ref{Eqn:MaterialNetwork}) is implemented 
as a ResNet~\cite{ResNet2}. Its input consists of $\mu''$ with $3 \times 5$
dimensions and the 7-dimensional shading parameters $P(x, \omega_r)$,
amounting to a total of 22 dimensions. The shading parameters include the
one-dimensional viewing direction, represented in tangent space as aforementioned,
and a six-dimensional material specification based on the Disney principled BSDF
parameters~\cite{ShadingFilmGames} -- specifically roughness, specularity, metallicity,
and RGB base color $\rho$.
The network itself is composed of six ResNet blocks, each containing 16 neurons
per layer. Bias terms are included in all layers, and fully connected layers at both
the input and output serve as dimension adapters for the network.
All activation functions are exponential linear units
\begin{equation} \label{Eqn:ELU}
  \text{\href{https://closeheat.com/blog/elu-activation-function}{ELU}}(x) := \begin{cases} x & x \geq 0\\
  \alpha (e^x - 1) & x < 0 \end{cases}
\end{equation}
with $\alpha = 1$. The network has about 3754 weights.

\begin{figure}
\centering
\begin{tabular}{cc}
\includegraphics[width=0.48\linewidth]{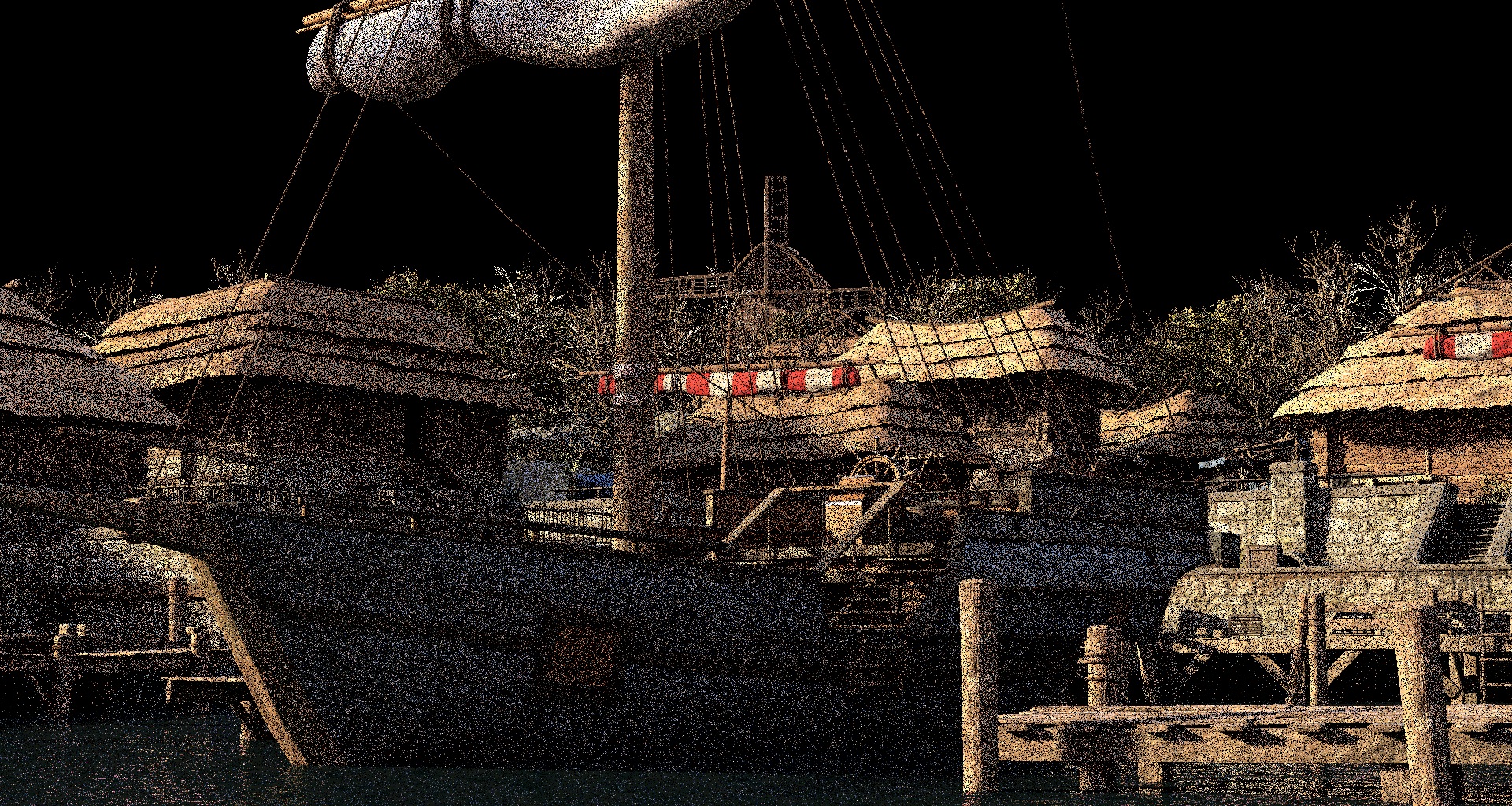}
& \includegraphics[width=0.48\linewidth]{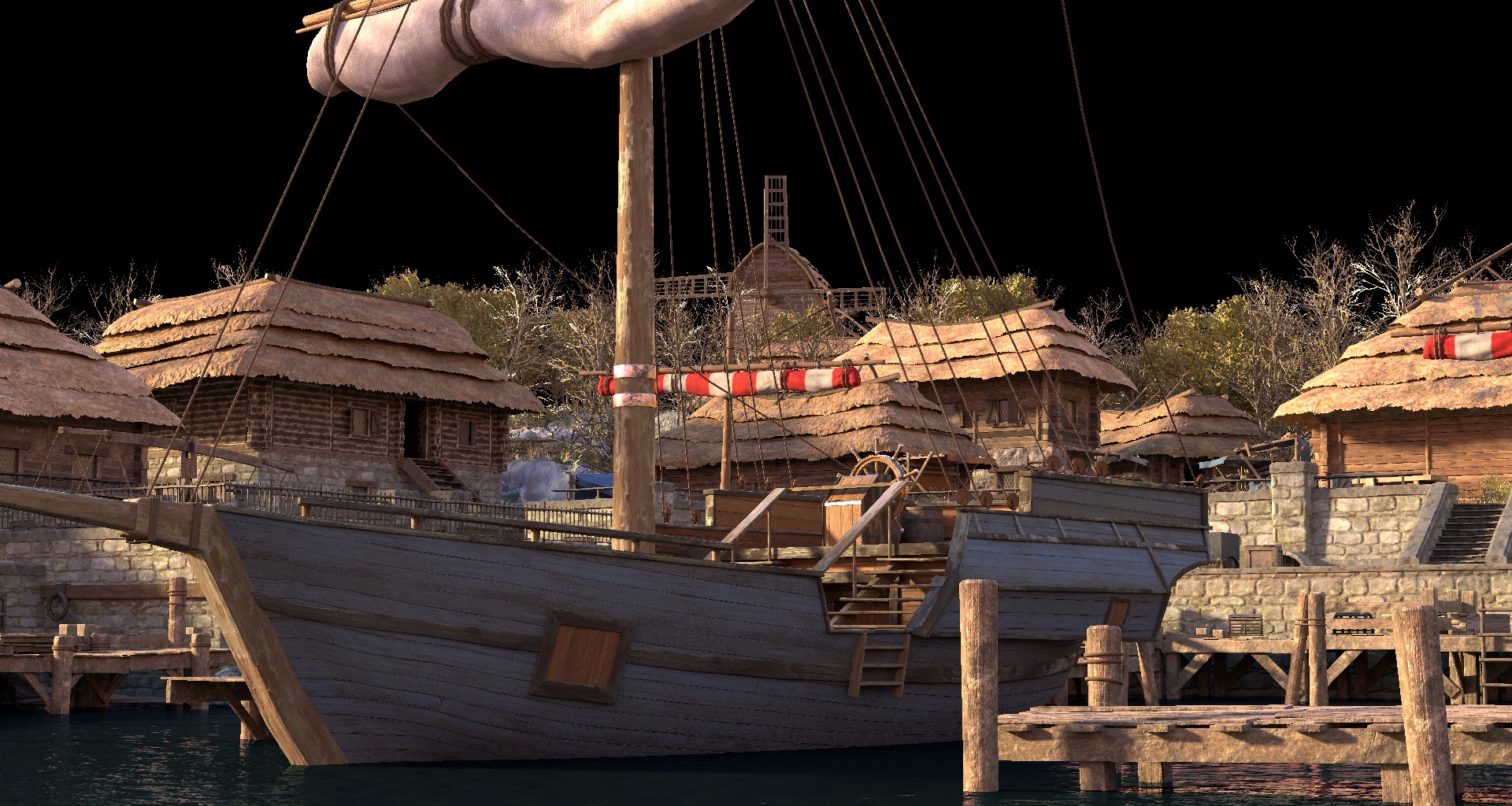} \\
a) $M_\theta(x, \omega_r, \mu)$, 1spp 
& $\overline{M}_\theta(x, \omega_r, \mu)$, 2048 spp \\ \\
& \includegraphics[width=0.48\linewidth]{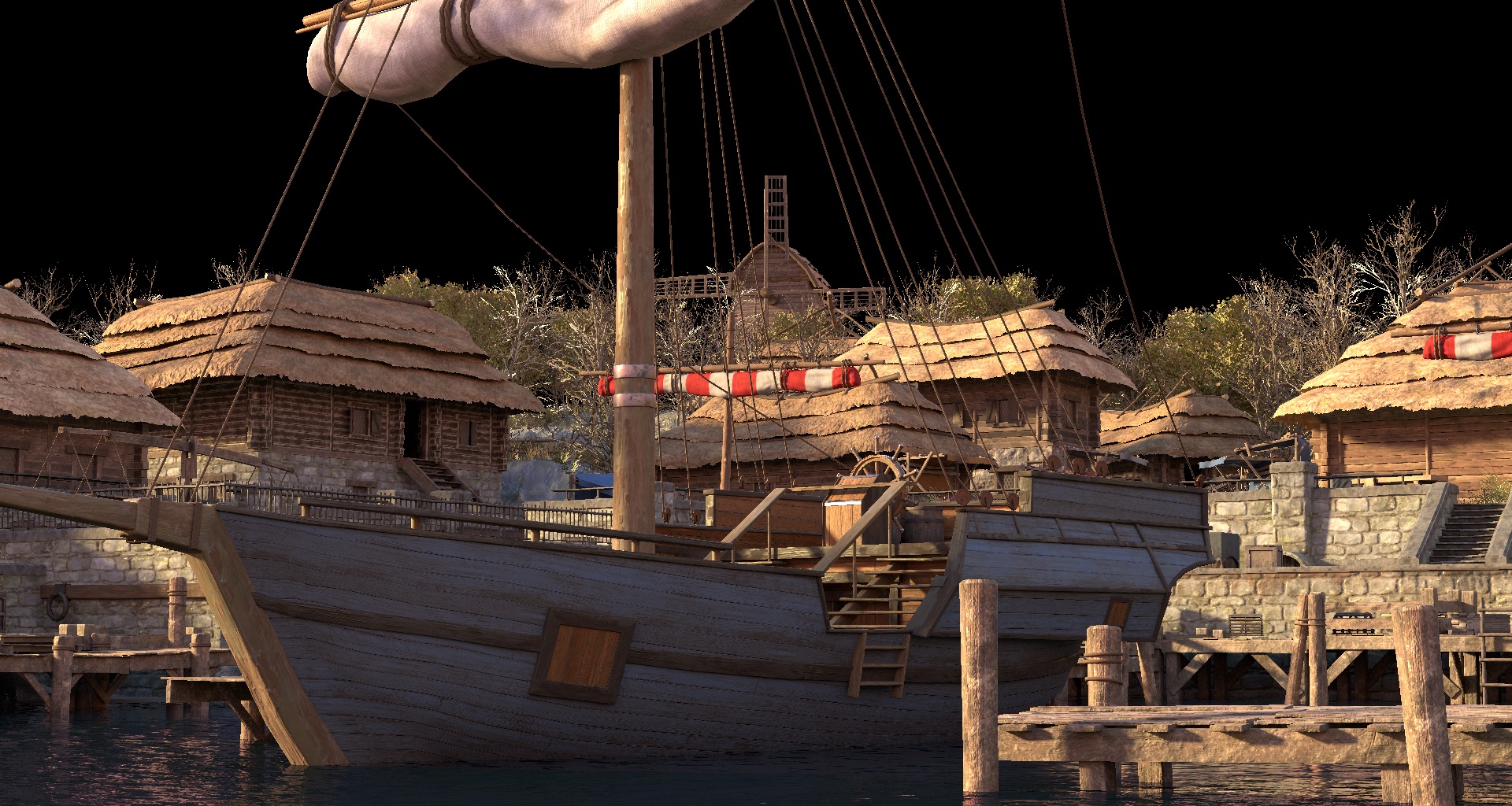} \\
& $M_\theta(x, \omega_r, \mu^*)$, 2048 spp \\ \\
\includegraphics[width=0.48\linewidth]{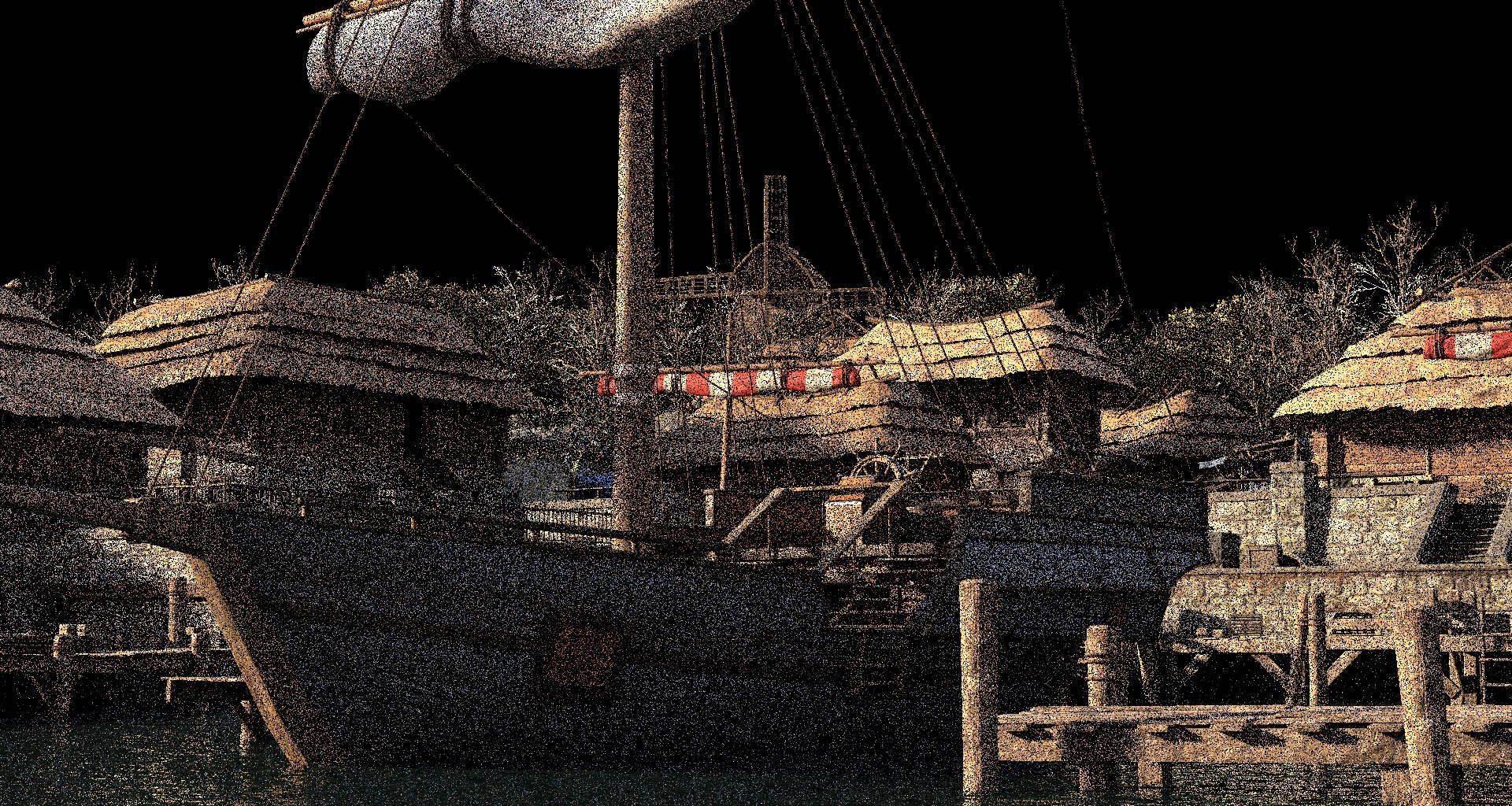}
& \includegraphics[width=0.48\linewidth]{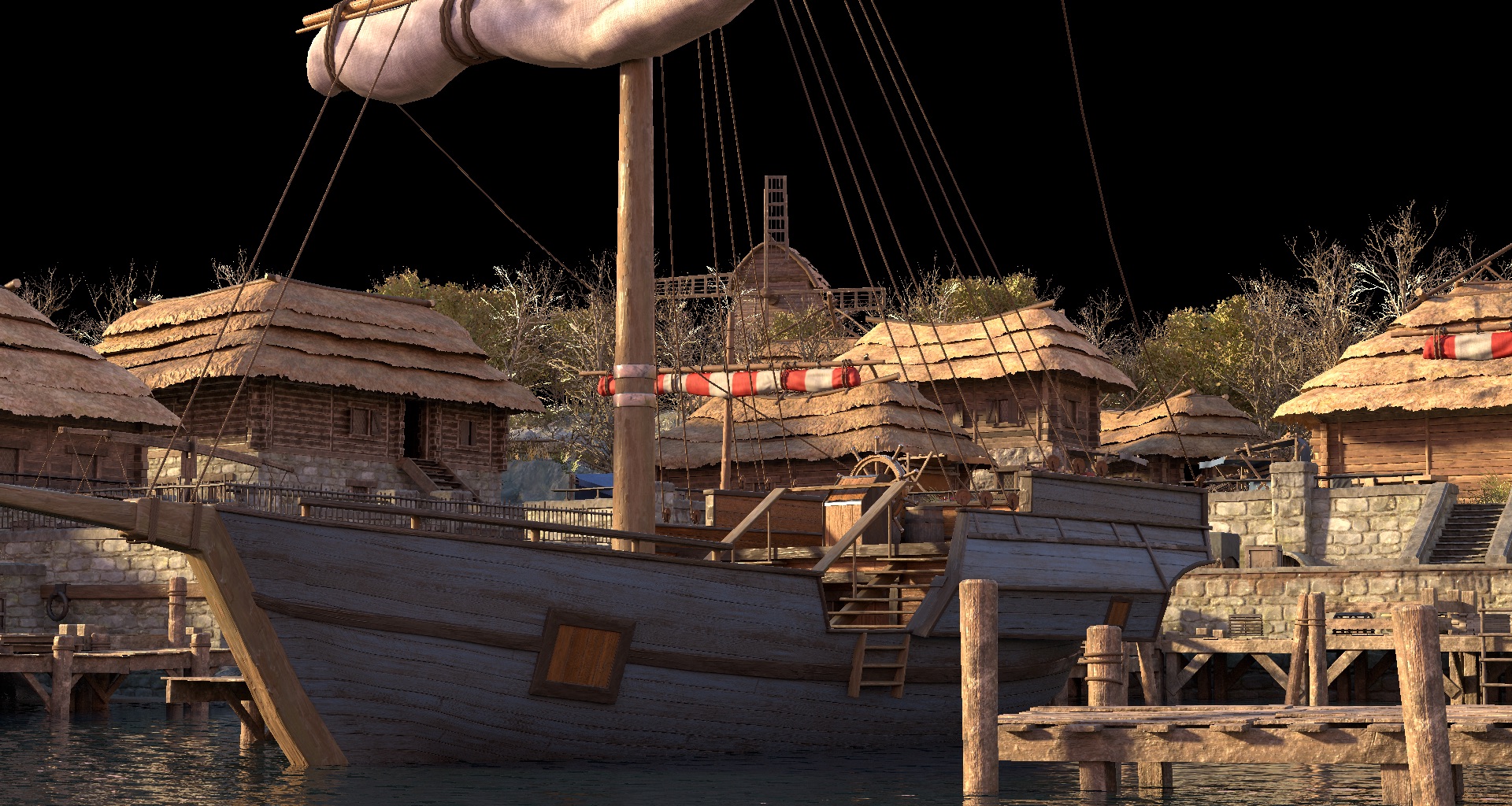} \\
b) Unreal Engine BSDF, 1spp & 2048 spp \\ \\
& \includegraphics[width=0.48\linewidth]{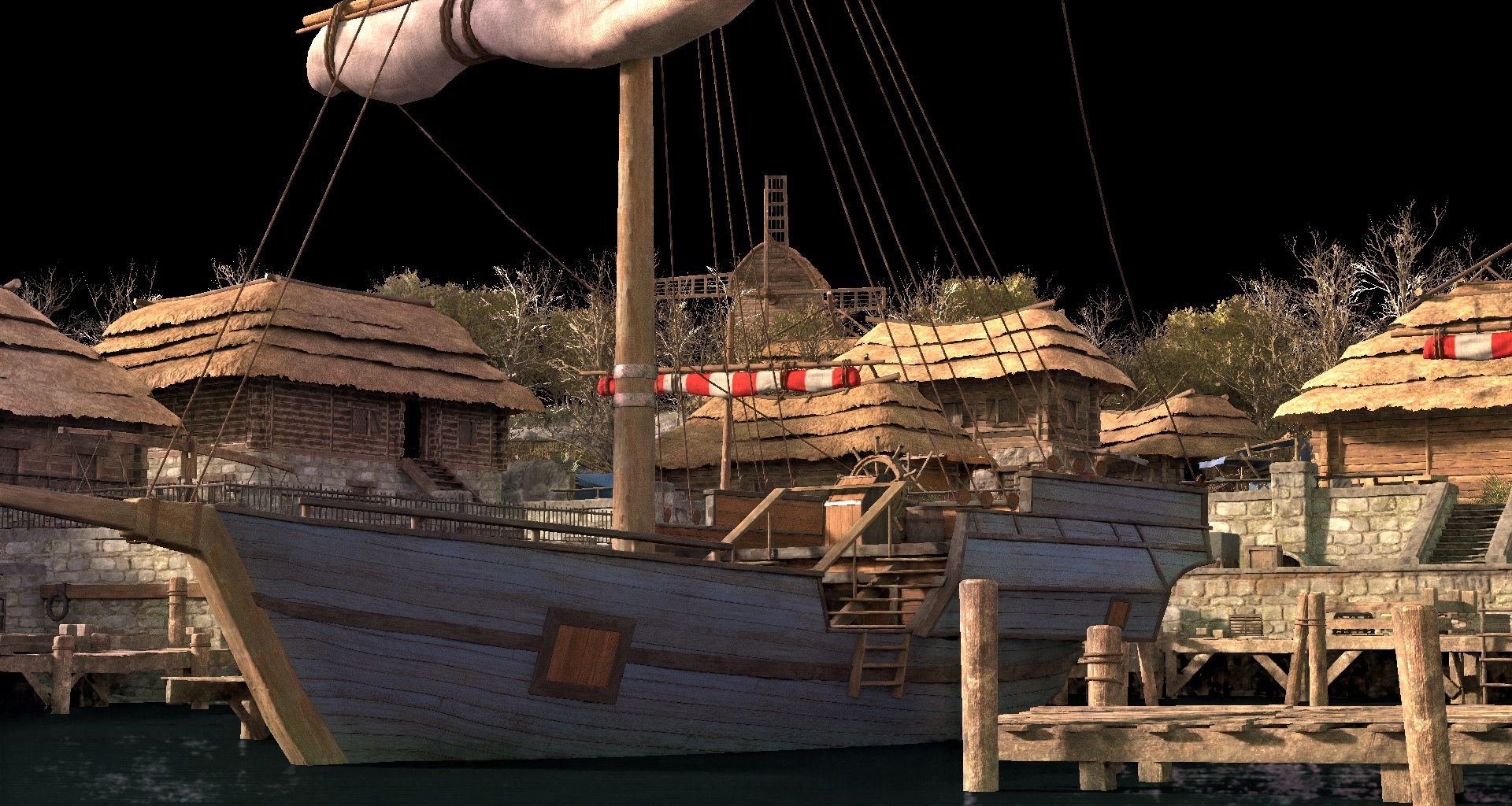} \\
& c) material agnostic denoising (MAD), 1spp
\end{tabular}
\caption{Comparison of
a) using the trained material decoder $M_\theta$ in (\ref{Eqn:Approximation}) for shading at 1 and averaged 2048 samples per pixel (spp),
and to the projected irradiance $\mu^*$, approximated by 2048 spp,
b) the reference as provided by the Unreal Engine BSDF, and
c) material agnostic denoising (MAD).
}
\label{Fig:NeuralMaterial}
\end{figure}

\subsection{Results} \label{Sec:Results}

After training the material decoder and denoiser (Sect.~\ref{Sec:Training}), the neural shading
pipeline (Sect.~\ref{Sec:Algorithm}) runs in real-time.
Fig.~\ref{Fig:NeuralMaterial} compares the original BSDF as used to compute (\ref{Eqn:IEQ})
to the result of the machine learned neural integral operator in (\ref{Eqn:Approximation}).

\paragraph{Avoiding Division by Zero.}

To avoid numerical issues caused by the division by small numbers $x$,
a too small divisor may be avoided by $\max\{x, \epsilon\}$ or
by adding the small number $\epsilon > 0$, i.e. dividing by $x + \epsilon$.
The former yields the actually desired result for $x > \epsilon$,
the latter smoothly compresses the signal towards zero and hence
always is a tiny bit off the desired fraction.
While the observable differences in normalization in (\ref{Eqn:Compress})
and (\ref{Eqn:Normalization2}) are subtle at least, restoring the scale in
(\ref{Eqn:Denoiser}) and (\ref{Eqn:Interpolation}) is best with the original
divisor without $\epsilon$.

\section{Training the Neural Networks} \label{Sec:Training}

The neural networks are trained in two separate passes:
First, the material
decoder $M_\theta$ is trained without the denoiser in the loop.
Then, the denoiser is trained using the already trained material decoder.
Note that in light transport simulation infinite sequences of
unique, unbiased training pairs can be generated on the fly just by sampling.

\subsection{Training the Material Decoder} \label{Sec:TrainDecoder}

The material decoder
$M_\theta$ is trained using procedurally generated examples
without any ray tracing.
The approximation (\ref{Eqn:Approximation})
only depends on $(x, \omega_r)$ and hence is independent of statistics
across pixels, which simplifies training.
Therefore, instead of evaluating the BSDF parameters in a specific location $x$,
material parameters are uniformly randomly sampled
from their respective ranges (Sect.~\ref{Sec:MaterialDecoder}) for each training example.

Four uniformly distributed directions of incident radiance are sampled,
alongside a uniformly randomly sampled outgoing direction $\omega_r$.
The reference is computed by evaluating the material implementation (Sect.~\ref{Sec:Disney}).
Each light sample has a random RGB radiance uniformly sampled within $[0, 16]^3$. Note that the range is
a free parameter since the intensity is decoupled from the neural network
inputs, as explained in Sect.~\ref{Sec:MaterialDecoder}.

\paragraph{Loss Function.}

The loss function is crucial for learning a visual
appearance that matches the original material. Using an $L^1$-loss function makes
the approximation work well in the darker regions of the BSDF, whereas the
$L^2$-loss focusses accuracy on the highlights. Instead, we are using a relative MSE based on the
pixel loss function
\[
  {\mathcal L}(x, \omega_r, \mu) = \frac{(M_\theta(x, \omega_r, \mu) - L_r(x, \omega_r))^2}{\max\{L_r^2(x, \omega_r), \epsilon\}}
\]
that computes an error that is relative to the magnitude of the function and hence maintains the
visual appearance across all regions of the BSDF.
The parameters for the optimizer are provided in Sect.~\ref{Sec:Parameters}.

\paragraph{Material simplifications:}

For specific choices of parameters, an analytic BSDF model may degenerate
to a Dirac delta function. Representing such functions accurately would require a very large
network capacity and hurt training convergence. We therefore both limit the
roughness of the material to a minimum of 0.1 and restrict the output
range of the BSDF to assume a maximum value of 16.0.

\paragraph{Importance sampling of highlights:}

The accuracy of the model depends on the training data set
and is improved by explicitly sampling respective parts of the BSDF model.
With a probability of
95\% we perform uniform sampling for the light source directions. For the
remaining samples, we use the importance sampling routine of the underlying BSDF to focus samples
onto the highlights (Sect.~\ref{Sec:Disney}). In the importance sampling we clamp the roughness to a
minimum of 0.5 which ensures good coverage over the entire highlight instead
of just the portions with the highest throughput.

\subsection{Dependent Training of the Denoiser} \label{Sec:Dependent}

Given the trained material decoder, the denoiser
convolutional U-Net \cite{U-Net} is trained noise-to-noise~\cite{Noise2Noise}
on randomly selected screen tiles. Since the range compression (\ref{Eqn:Compress})
is based on a wide blur, we use large tiles of $512\times512$ pixels,
which also ensures that the boundary handling of the tiles does not dominate the
gradients.
Using real scene geometry and one sample per pixel, the locations $x$,
the normals $\hat n$, and the depth buffer $d$ are rendered. For each pixel
BSDF parameters are sampled uniformly in
order to train independently from the actual scene materials. 
Sharing this data, two independently sampled projected space representations
$\mu_A$ und $\mu_B$ (Sect.~\ref{Sec:IrradianceEncoder}) are generated
that obviously exhibit identically distributed noise level statistics across the pixels.
A random position is sampled on the light source and visibility is computed using ray
tracing.
These training examples are shaded using the neural BSDF, instead of the reference BSDF. This
way differences between the neural BSDF and reference BSDF do
not influence the training process.
Generating the training set takes about a minute and 16GB of memory.

\paragraph{Loss Functions.}

Our primary loss function is a relative MSE loss function such that each
training example and each region of a training example is uniformly
weighted in the optimization procedure. As in the
Noise2Noise \cite{Noise2Noise} scenario the reference is not available for normalizing the
loss, the loss is
normalized by the denoised result. By disabling (\texttt{detach}) the gradient flow
for the denominator of the loss, the loss remains unbiased:
\[
  {\mathcal L}(\mu^c_A, \mu^c_B)
  = \frac{(M_\theta(U_\theta(\mu^c_A)) - M_\theta(\mu^c_B))^2}{\max\{0.5 * M_\theta(U^\text{\tt detach}_\theta(\mu^c_A))^2 + 0.5 * M_\theta(U^\text{\tt detach}_\theta(\mu^c_B))^2,\epsilon\}}
\]
where $\epsilon = 0.00001$. The superscript $c$ is the shorthand for the compressed
projected irradiance as in (\ref{Eqn:Compress}). As reported by Lehtinen et al. \cite{Noise2Noise},
the stability of optimization is improved by normalizing by the average of the squares of the
denoiser outputs both for the input and the reference. Furthermore, we clamp the
minimum allowed value instead of the summation as described in
\cite{Noise2Noise} since this otherwise warps the normalization.

We propose an additional loss that we call consistency constraint that is
particularly helpful for low numbers of samples per pixel (spp) in Noise2Noise training. The concept of the loss
function is that for two realizations of samples for the same frame, the denoiser
should be producing identical output. We use the same relative MSE loss as
above, however the loss now measures the error between the two denoised images:
\[
  {\mathcal L}(\mu^c_A, \mu^c_B)
  = \frac{(M_\theta(U_\theta(\mu^c_A)) - M_\theta(U^\text{\tt detach}_\theta(\mu^c_B)))^2}{\max\{0.5 * M_\theta(U^\text{\tt detach}_\theta(\mu^c_A))^2 + 0.5 * M_\theta(U^\text{\tt detach}_\theta(\mu^c_B))^2,\epsilon\}}
\]
As before, the gradient flow is only enabled for the prediction. Swapping the
roles of the $\mu_A$ used for prediction and $\mu_B$ used for the reference
provide an efficient data augmentation.
The parameters for the optimizer are provided in Sect.~\ref{Sec:Parameters}.

Theoretically, we may fully randomize the material parameters that are passed
to the shading network instead of sharing the data per pixel.
This would
have the benefit of increasing the training data. However, we
found that this negatively impacts the training results.
Similarly,
formulating the loss function in the projected space instead of
evaluating the full shading function with a loss formulated in RGB
did not yield usable results. The training is augmented by randomly
permuting the RGB channels and randomly zeroing out one color channel
in order to prevent the denoiser from mixing colors across channels
and yet taking advantage of the shared geometry information.

\subsection{Independent Training of the Denoiser}

Alternatively, the denoiser can be trained Noise2Noise independent of the
material decoder. The simplified training procedure then just independently
samples two projected space representations
$\mu_A$ und $\mu_B$ (Sect.~\ref{Sec:IrradianceEncoder})
as in the previous section.

Then, the relative MSE loss function amounts to
\[
  {\mathcal L}(\mu^c_A, \mu^c_B)
  = \frac{(U_\theta(\mu^c_A) - \mu^c_B)^2}{\max\{0.5 * U^\text{\tt detach}_\theta(\mu^c_A)^2 + 0.5 * U^\text{\tt detach}_\theta(\mu^c_B)^2,\epsilon\}}
\]
and the consistency constraint becomes
\[
  {\mathcal L}(\mu^c_A, \mu^c_B)
  = \frac{(U_\theta(\mu^c_A) - U^\text{\tt detach}_\theta(\mu^c_B))^2}{\max\{0.5 * U^\text{\tt detach}_\theta(\mu^c_A)^2 + 0.5 * U^\text{\tt detach}_\theta(\mu^c_B)^2,\epsilon\}} .
\]

\section{Discussion} \label{Sec:Discussion}

The idea of filtering irradiance in world space before shading and
the higher-dimensional structure of irradiance gradients has been
received early \cite{WH:92}. On the one hand, an entire body of work \cite{Krivanek:2009:PGI}
culminates in the neural radiance cache \cite{muller2021nrc} that
stores radiance using a multiresolution hash encoding \cite{mueller2022instant}
and trains in real-time using temporal accumulation across frames.
On the other hand, the neural shading pipeline requires data from only a single frame
and overcomes imperfectly hand-crafted edge stopping
heuristics \cite{WH:92,KellerPhD,ATrous,PathSpaceFiltering,HashedPSF}
simply by machine learning.

As synthetic data generation (SDG) is intrinsic to light transport simulation,
it is ideal to explore neural network architectures for
approximating operators \cite{DeepONet,FNO,AFNO}.
We hence found that the U-Net $U_\theta$ (see Fig.~\ref{Fig:U-Net}) used for denoising in (\ref{Eqn:Denoiser})
may be realized using only the rectified linear unit (ReLU) activation function
$\sigma(x) := \max\{0, x\}$ without any bias terms and normalizations.
Such a U-Net is invariant to scaling the input by a positive real number $\alpha \in \mathbb{R}_0^+$, because
$\sigma(\alpha x) = \max\{0, \alpha x\} = \alpha \sigma(x)$.

While not using any normalization, the material decoder requires bias terms.
Furthermore, the continuously differentiable ELU activation function in (\ref{Eqn:ELU}) reduces
visible segmentation artifacts due geometric discontinuities
that the discontinuous ReLU activation function does not hide well.

\subsection{Subsampling}

The algorithm is amenable to subsampling. For example, light transport paths may
be sampled once in every 2x2 pixel block. The denoiser then fills in the information
for the pixels not sampled before shading.

\subsection{Temporal Anti-Aliasing (TAA)}

As the neural shading pipeline works with any sampling method
and both rasterization and ray tracing,
temporal anti-aliasing is straightforward to add.
Interestingly, scintillation artifacts across frames are observed when sampling with real-time
algorithms like ReSTIR \cite{CourseReSTIR}, whose efficiency relies on temporally
correlated path reuse. The phenomenon is known since
the early days of irradiance interpolation \cite{Krivanek:2009:PGI} and still subject of active research. To a certain extend this is
to be expected, when the denoiser does not perfectly reconstruct the brightness
of the parametric integral (\ref{Eqn:IEQ}) due to extremely low sampling rates.

Scintillation is caused
by rare event bright samples, too. These so-called fireflies offset averages a lot, but
appear rarely and incoherently across frames and pixels, causing transient temporal artifacts.
This is still an issue of sampling algorithms that only partially can be
handled by multiple importance sampling. In practice, samples
may be clipped by a firefly filter to attenuate the artifacts.

\subsection{Future Work}

Our algorithm does neither consider depth of field nor motion blur,
and is restricted to isotropic BSDFs. In fact, our material decoder is trained
to match one specific BSDF model. Anisotropic BSDFs
require the rotation of BSDF lobes and potentially an extension of the
projected space representation. The extension to support
multiple BSDF models and shader baking by multiple sets of weights for the material
decoder is an avenue of future work.

As the material decoder is trained independently from scene geometry and separately
from the denoiser, there is an opportunity
to explore the feasibility of real-time training the material
decoder as a cache for complex appearances.
Furthermore, the parameters for the material decoder may be provided by
neural textures \cite{NeuralTexture}, and
the material decoder itself may be trained to represent appearance \cite{NeuralAppearance}.

Our method works with any noise filter
and while we train a U-Net for that purpose, there is ample opportunity
for optimization by replacing at least parts of the neural network filters
by analytic components~\cite{AnalyticConvolutions}.
In a similar way, the parameters of the linear material decoder model in (\ref{Eqn:Interpolation}) may
be directly provided by the neural network $m_\theta$ without the need of the softmax-normalization in (\ref{Eqn:Weights}).

\section{Conclusion}

We introduce the neural shading pipeline as an efficient
real-time rendering method for extremely low sampling
rates. Our algorithm is straightforward to implement,
operating on data from only a single frame without the need
for temporal information such as motion vectors. 
Additionally, we propose an efficient training procedure
that is independent of the actual material parameters of the content to
be rendered. Beyond computer graphics,
we expect this approach to be applicable to the general
computation of functionals of the solution of integral equations.

\section*{Dedication}

This article is dedicated to Jerome Spanier, the pioneering gentleman of MCQMC.


\appendix
\section{Appendix}

\subsection{Disney Principled BSDF Model} \label{Sec:Disney}

For the integral kernel $f$, we employ the isotropic Disney principled
bidirectional scattering distribution function (BSDF) \cite{ShadingFilmGames}
without clear coat as implemented by the \texttt{DefaultLitMaterial} in the Unreal Engine 4.
For a white base color (=1), we select the BSDF parameters as 
\begin{center}
\begin{tabular}{c|c|c|c}
& metallic & specular & roughness \\ \hline
$E_1$ & 0.5 & 0.5 & 0.1 \\
$E_2$ & 1.0 & 0.0 & 0.1 \\
$E_3$ & 0.0 & 1.0 & 0.1 \\
$E_4$ & 1.0 & 1.0 & 0.6
\end{tabular}
\end{center}
and $E_0(\omega, \omega_r) := 1$, amounting to $d = 5$, to define the components (Sect.~\ref{Sec:IrradianceEncoder}) of
\begin{eqnarray*}
    \bsE(\omega, \omega_r) & := &  \left(1, \raisebox{-2ex}{\includegraphics[width=0.1\linewidth]{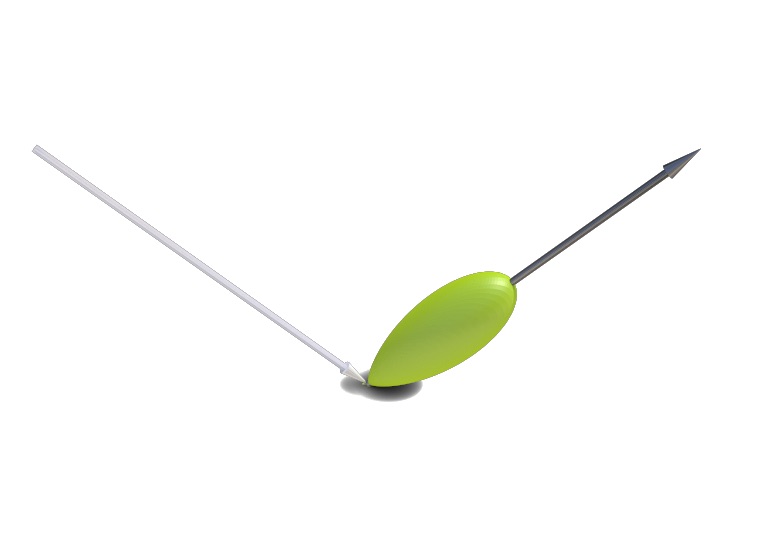}},
    \raisebox{-2ex}{\includegraphics[width=0.1\linewidth]{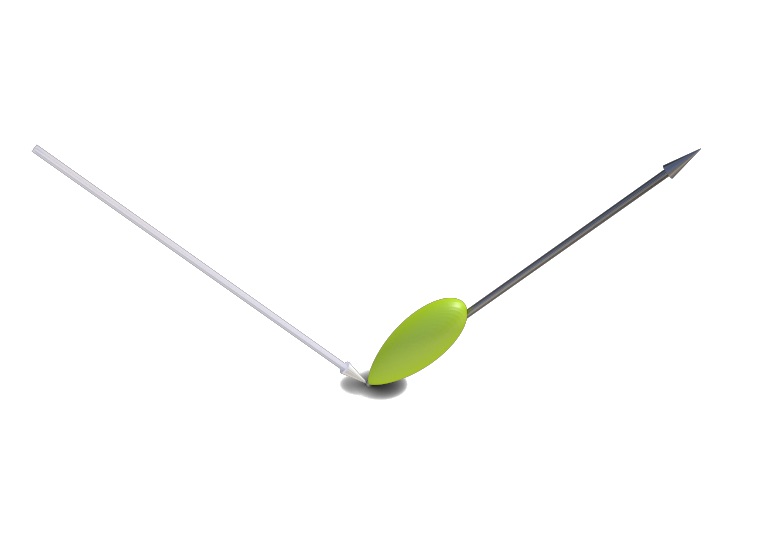}}, \ldots \right) 
    \text{,  where, for illustration, } \\
     f(x, \omega, \omega_r)
   & = & \rho_d(x) \cdot \raisebox{-2ex}{\includegraphics[width=0.1\linewidth]{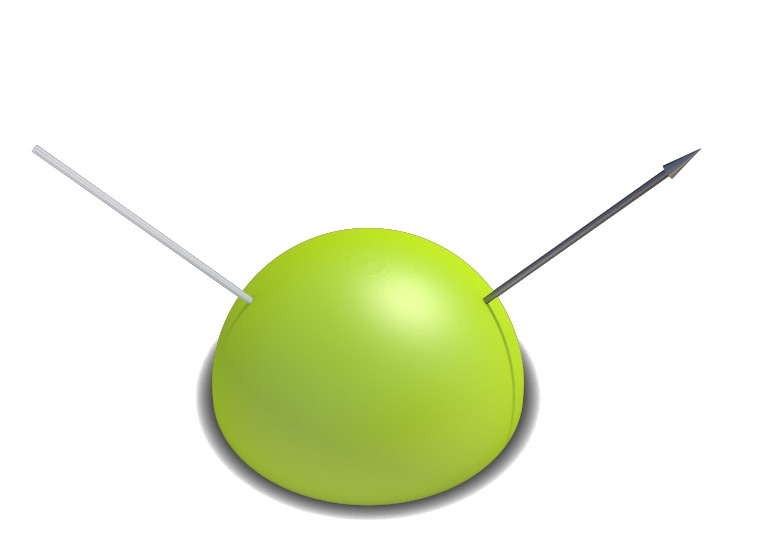}}
   + \rho_g(x) \cdot \raisebox{-2ex}{\includegraphics[width=0.1\linewidth]{Phong_2_specular}}
   + \rho_s(x) \cdot \raisebox{-2ex}{\includegraphics[width=0.1\linewidth]{Phong_2_specularCos}} .
\end{eqnarray*}

\subsection{Training Parameters} \label{Sec:Parameters}

We use the AdamW optimizer with the PyTorch default parameters \texttt{(betas = (0.9,
0.999), eps = 1e-08, weight\_decay = 0.01, amsgrad = False)}.

The material decoder in Sect.~\ref{Sec:TrainDecoder} is trained with
a learning
rate of \texttt{1e-3} and a batch size of $2^{18}$. For training stability we
employ clamping of the gradient $L^2$-norm to a maximum of 1. This helps with
training convergence and avoids situations where exploding gradients would
derail the optimization process.
The denoiser in Sect.~\ref{Sec:Dependent} is trained with a learning
rate of \texttt{1e-3} and a batch size of 4.

\end{document}